\documentclass[twocolumn]{aastex62}

\newcommand{\teff}{\mbox{T$_{\rm eff}$}}
\newcommand{\logg}{\mbox{log~{\it g}}}
\newcommand{\vmicro}{\mbox{$\xi_{\rm t}$}}
\newcommand{\kmsec}{\mbox{km~s$^{\rm -1}$}}

\newcommand{\x}{\mbox{$\Delta_{\tiny{\mathrm{F275W,F814W}}}$}}
\newcommand{\y}{\mbox{$\Delta_{\tiny{C~\mathrm{ F275W,F336W,F438W}}}$}}

\accepted{October 4, 2019}
\submitjournal{ApJ}

\shorttitle{Chemical abundances in NGC\,3201} 
\shortauthors{A.\,F. Marino, et al.} 

\begin{document}

\title{Chemical abundances along the 1G sequence of the chromosome maps:
  The Globular Cluster NGC\,3201\footnote{
Based on observations collected at the European Southern Observatory
under ESO programme 0101.D-0113(A), and the NASA/ESA {\it Hubble Space
  Telescope}, obtained at the Space Telescope Science Institute, which
is operated by AURA, Inc., under NASA contract NAS 5-26555. }}

\author{A.\ F.\,Marino} 
\affiliation{Dipartimento di Fisica e Astronomia ``Galileo Galilei'' - Univ. di Padova, Vicolo dell'Osservatorio 3, Padova, IT-35122}
\affiliation{Centro di Ateneo di Studi e Attivita' Spaziali ``Giuseppe Colombo'' - CISAS, Via Venezia 15, Padova, IT-35131}  
\author{A.\ P.\,Milone}
\affiliation{Dipartimento di Fisica e Astronomia ``Galileo Galilei'' - Univ. di Padova, Vicolo dell'Osservatorio 3, Padova, IT-35122}
\author{A.\ Sills}
\affiliation{Department of Physics \& Astronomy, McMaster University, 1280 Main Street West, Hamilton, ON, L8S 4M1, CANADA}
\author{D.\ Yong}
\affiliation{Research School of Astronomy \& Astrophysics, Australian National University, Canberra, ACT 2611, Australia} 
\author{A.\ Renzini}
\affiliation{Istituto Nazionale di Astrofisica - Osservatorio Astronomico di Padova, Vicolo dell'Osservatorio 5, Padova, IT-35122} 
\author{L.\ R.\,Bedin}
\affiliation{Istituto Nazionale di Astrofisica - Osservatorio Astronomico di Padova, Vicolo dell'Osservatorio 5, Padova, IT-35122}
\author{G.\ Cordoni}
\affiliation{Dipartimento di Fisica e Astronomia ``Galileo Galilei'' - Univ. di Padova, Vicolo dell'Osservatorio 3, Padova, IT-35122}
\author{F.\ D'Antona}
\affiliation{Istituto Nazionale di Astrofisica - Osservatorio Astronomico di Roma, Via Frascati 33, I-00040 Monteporzio Catone, Roma, Italy}
\author{H.\ Jerjen}
\affiliation{Research School of Astronomy \& Astrophysics, Australian National University, Canberra, ACT 2611, Australia}  
\author{A.\ Karakas}
\affiliation{School of Physics \& Astronomy, Monash University, Clayton 3800, Victoria, Australia}  
\author{E.\ Lagioia}
\affiliation{Dipartimento di Fisica e Astronomia ``Galileo Galilei'' - Univ. di Padova, Vicolo dell'Osservatorio 3, Padova, IT-35122}
\author{G.\ Piotto}
\affiliation{Dipartimento di Fisica e Astronomia ``Galileo Galilei'' - Univ. di Padova, Vicolo dell'Osservatorio 3, Padova, IT-35122}
\author{M.\ Tailo}
\affiliation{Dipartimento di Fisica e Astronomia ``Galileo Galilei'' - Univ. di Padova, Vicolo dell'Osservatorio 3, Padova, IT-35122}

\correspondingauthor{A.\ F.\,Marino}
\email{anna.marino@unipd.it}

\begin{abstract}
The {\it Hubble~Space~Telescope} ($HST$) UV Legacy Survey of Galactic Globular
Clusters (GCs) has investigated multiple stellar populations by
means of the ``chromosome map'' (ChM) diagnostic tool that maximises the separation
between stars with different chemical composition. 
One of the most challenging features revealed by ChMs analysis is the apparent
inhomogeneity among stars belonging to the first
population, a phenomenon largely attributed to He variations. 
However, this explanation is not supported by the uniformity in $p$-capture
elements of these stars.
The $HST$ survey has revealed that the GC NGC\,3201 shows an
exceptionally wide coverage in the \x\ parameter of the ChM.
We present a chemical abundance analysis of 24 elements in 18 giants
belonging to the first population of this GC, 
and having a wide range in \x. 
As far as the $p$-capture elements are concerned, the chemical
abundances are typical of 1G stars, as expected from the
location of our targets in the ChM. 
Based on radial velocities and
chemical abundances arguments, we find that the three stars
with the lowest \x\ values are binary candidates. This suggests that,
at least those stars could be explained with binarity.
These results are consistent with evidence inferred from
multi-band photometry that evolved blue stragglers populate the
bluest part of the 1G sequence in the ChM.
The remaining 15 spectroscopic targets show a small range in the
overall metallicity by $\sim$0.10~dex, 
with stars at higher \x\ values having higher absolute abundances.
We suggest that a small variation in metals and binarity
govern the color spread of the 1G in the ChM, and that evolved blue
stragglers contribute to the bluest tail of the 1G sequence.    
\end{abstract}


\section{Introduction}\label{sec:intro}

The presence of multiple stellar populations in globular clusters
(GCs) is a well-assessed fact. 
Chemical abundance variations within stars in a given GC have
been known for many years now, with some of the earliest studies
including \citet{Popper}, \citet{Harding}, \citet{Osborn},
\citet{Cohen}. More recently, different stellar populations have been 
detected through multiple sequences along the color-magnitude diagram
\citep[CMD, e.g.\,][]{APM12}. The photometrically-observed multiple
sequences are associated with variations in chemical elements involved in hot-H
burning \citep[light elements, CNONa, see\,][for the earliest studies
on this issue]{AFM08, Yong08a}. One stellar
population, usually considered the first population (or generation),
is chemically similar to field halo stars,  
while the other populations display various degrees of enrichment in
He/N/Na, depletion in C, and
depletion of O with respect to the first population,
as opposite to the $\alpha$-enhancement typical in halo field stars
\citep[e.g.\,][]{Carretta09, Gratton12}.  

On top of this typical observed pattern, the recent analysis of
UV high-precision data from the {\it Hubble Space Telescope} ($HST$) has
revealed an even more complex picture \citep[][]{Mil15, Mil17}. 
All the GCs exhibit two main discrete groups of first-generation (1G) and
second-generation (2G) stars along the ``chromosome
map'' (ChMs), that constitute the most effective and successful
diagnostic tool to isolate the different populations of stars hosted in a GC.
These sort of two-colour photometric diagrams,
constructed by combining multi-filter $HST$ images in $m_{\rm F275W}$,
$m_{\rm F336W}$, $m_{\rm F438W}$, $m_{\rm F814W}$, are highly sensitive
to the chemistry of the different stellar populations. 
On a typical ChM plane, \y\ vs.\,\x\footnote{See
  Section~\ref{sec:data}. A detailed definition of \x\ and \y\ can be
  found in \citet{Mil15, Mil17}.}, 
all the observed GCs display a typical ChM shape, with a
different number of seemingly-discrete groups, appearing along the main
pattern, that are associated with distinct stellar populations. 
By combining spectroscopy and photometry we find that 1G 
stars on the ChMs, with low \y\ values, share the same chemical composition of
halo field stars at the same metallicity, while 2G stars, are enhanced
in He, N and Na and depleted in C and O \citep{Mar19}. 

The ChMs exhibit a variety of morphologies both in terms of
number of stellar populations and in terms of shape and
extension. \citet{Mil17} subdivided 57 analysed maps in two
main groups: (i) Type~I GCs having a single ChM pattern, with a 1G and
two or more 2G populations, and (ii) Type~II GCs displaying multiple
ChMs, with minor populations located on red additional ChMs. 
While we refer to \citet{Mar19} for a
detailed chemical charaterisation of these two main classes of maps, here we just recall that the
additional red stellar populations in Type~II GCs are
enhanced in metallicity, and (in some cases) in the elements produced
via $slow$-neutron capture reactions \citep[$s$-elements,
e.g.\,][]{Yong08b, AFM09, AFM15}.

ChMs, besides being a powerful tool to explore the multiple stellar
population phenomenon, provide fundamental information that can shed
light on how GCs formed and evolved, which might be hidden in the 1G
stars themselves. 
Indeed, one of the most interesting features
observed on the ChMs is perhaps the apparent chemical inhomogeneity
within the 1G population.  
Indeed the 1G sequence on the ChM is either
elongated or bimodal, thus indicating that its stars are not
consistent with a simple stellar population. 
\citet{Mil15, Mil18} and \citet{Dantona16} suggested that
either a relatively large 
variation in He, even by $\Delta$Y$\sim$0.10, or a small variation in
metallicity, by $\sim$0.10~dex, within 1G stars could reproduce the
morphology of the 1G sequence on the map. The homogeneity in the
light elements among these stars observed in the ChM of many GCs 
\citep[][see also, \citealt{CZ19}]{Mar19}
suggests that, if He is responsible for the elongated distribution of 1G stars on the ChM, a
nucleosynthesis process occurred that changed the He content but left
untouched the elements involved in the standard H-burning, such as C and N
(see discussion in \citealt{Mil18}). As we
are not aware of the existence of such a mechanism, a small variation in
metallicity among 1G stars might be a more plausible solution \citep[][]{Mar19}.

An internal variation in helium among 1G stars seems not be supported
by recent analysis of horizontal branch stars in M\,3 \citep{Tailo19}. 
The fact that the first population on the maps is inconsistent with
one single group poses serious challenges to our current understanding
of multiple stellar populations. The possibility that 1G stars have an
internal spread in the overall metallicity would suggest that SNe played
a role in the star-formation history of most GCs and that these
stellar systems have been able to retain a small amount of material
polluted by the SNe. In such a case, either a fraction of the 1G could be
formed from material polluted from a previous generation of stars thus
indicating that only a fraction of 1G stars constitute the real
primordial stellar generation, making the mass-budget problem even
more dramatic; or the interstellar medium from which 1G stars formed
was inhomogeneous.
In contrast, the possibility that the 1G has an internal spread in helium
(but constant C, N, O, and Fe) could be indicative of some
still-unknown mechanism that occurs in GC stars 
only, possibly an accretion phenomenon in the pre-main sequence
phase.

In this work we provide a high-resolution spectroscopic
abundance analysis of 1G stars, as selected from the ChM, in the GC
NGC\,3201. 
This cluster, displaying a well-elongated 1G population on the ChM,
is an ideal target for an in-depth investigation of the chemical abundance
pattern within 1G stars. 
Although its ChM does not show evidence of additional stellar
populations on the red side, as in Type~II GCs, \citet{Dias18}
found multiple (anti)correlations in light elements, similar to those
observed in some of the Type~II clusters, such as M\,22 \citep{AFM09}. 
The layout of this paper is as follows:
Section~\ref{sec:data} presents the photometric and spectroscopic
data; Section~\ref{sec:atm} describes how we derive atmospheric
parameters; the chemical abundance analysis is presented in
Section~\ref{sec:abundances}; Sections~\ref{sec:results},
\ref{sec:binarity}, and \ref{sec:iron} discuss our results, that are
summarised in Section~\ref{sec:conclusions}.

\section{Data}\label{sec:data}

\subsection{The photometric dataset: the chromosome map of NGC\,3201 \label{sec:phot_data}} 

The photometric data used in this study come from the $HST$ UV Legacy
Survey which investigated multiple stellar populations in GCs
\citep[GO-13297,][]{Piotto15}.
Details on the images analyzed and on the data reduction can be found in
\citet{Piotto15} and \citet{Mil17}.
Photometry has been corrected for differential reddening effects as in
\citet{APM12bin}.

\citet{Mil17} analyzed the chromosome maps of 57 GCs. 
While we refer to \citet{Mil15, Mil17} for a detailed discussion on
how to construct these photometric diagrams, for the convenience of
the reader, we restate here their equations (1) and (2). Hence, \x\
and \y\ are defined as: 
\begin{equation}
\scriptsize
\x = W_{\rm {F275W,F814W}} \frac{X-X_{\rm {fiducial~R}}}{X_{\rm {fiducial~R}}-X_{\rm {fiducial~B}}}
\end{equation}
\begin{equation}
\scriptsize
\y = W_{\rm {C~F275W,F336W,F438W}} \frac{Y_{\rm {fiducial~R}}-Y}{Y_{\rm {fiducial~R}}-Y_{\rm {fiducial~B}}}
\end{equation}
where $X=(m_{\rm {F275W}}-m_{\rm {F814W}})$,  
$Y=C_{\rm {F275W,F336W,F438W}}$, with the latter being an index sensitive to
N abundances, and 'fiducial R' and 'fiducial B' correspond to the red
and the blue fiducial lines involved in the construction of the ChM. 

\begin{figure}
   \centering
   \includegraphics[width=0.47\textwidth]{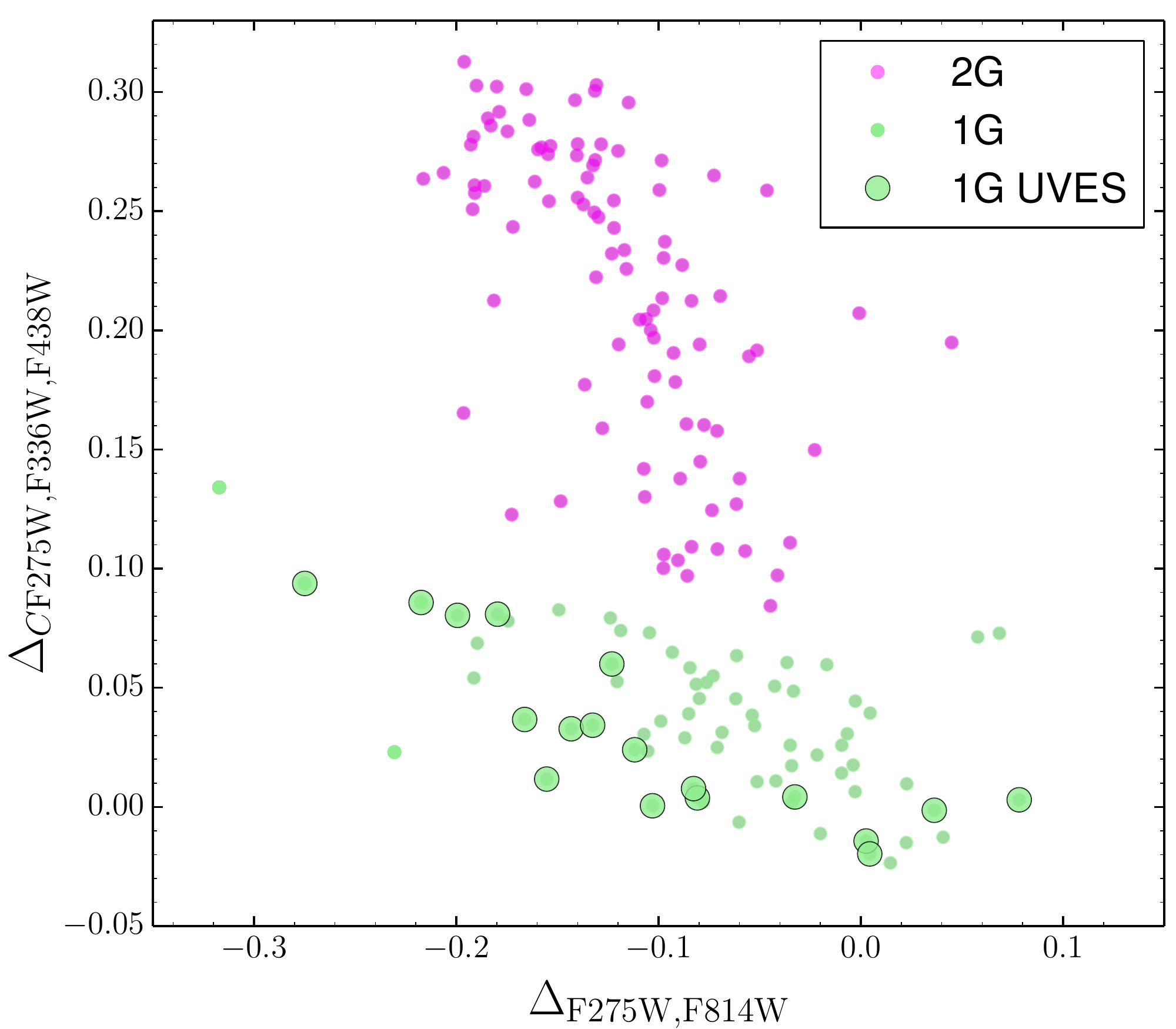}
      \caption{The chromosome map of NGC\,3201 from \citet{Mil17}. Stars in green and
      magenta represent 1G and 2G stars, respectively, as defined by Milone and
      co-workers. Our 18 spectroscopic UVES targets are all
      located in the 1G region, and are highlighted as larger filled circles.} 
        \label{fig:targets}
   \end{figure}

The analysis of the NGC\,3201 ChM revealed that this cluster covers a large range along
the \x\ axis, which is not consistent with the assumed chemical
homogeneity of the 1G (see their Fig.~4 and our Fig.~\ref{fig:targets}). 

\subsection{The spectroscopic dataset}\label{sec:spec_data}

Our spectroscopic data have been acquired using the FLAMES Ultraviolet and
Visual Echelle Spectrograph \citep[FLAMES-UVES,][]{Pasquini00} on the
European Southern Observatory’s (ESO) Very Large 
Telescope (VLT), through the program 0101.D-0113(A). 
The observations were taken in the standard RED580 setup, which has a
wavelength coverage of 4726-6835~\AA\ and a resolution $R \sim$47,000 \citep{Dekker00}.

Spectra are based on 5$\times$2775s exposures for the seven brightest stars
in our sample with $V$ mag between $\sim$13.5 and $\sim$14.5, and
23$\times$2775s for the 
three faintest ones with $V>$15. Stars with intermediate mag have
11$\times$2775s or 12$\times$2775s exposures on them (see Table~1).  
Data were reduced using the FLAMES-UVES pipelines within the
EsoReflex interface\footnote{\sf
  {https://www.eso.org/sci/software/esoreflex/}} \citep{Ballester00}, 
including bias subtraction, flat-field correction, wavelength
calibration, sky subtraction and spectral rectification. 
Once individual spectra were reduced, the telluric subtraction has
been performed by using the ESO MOLECFIT tool \citep[][]{smette14, kausch14}.

Radial velocities (RVs) were derived using
the {\sc iraf@FXCOR} task, which cross-correlates the object spectrum with a
template. For the template we used a synthetic spectrum obtained
through MOOG\footnote{{\sf http://www.as.utexas.edu/~chris/moog.html}} 
\citep[version June 2014,][]{moog}, computed with a model
stellar atmosphere interpolated from the 
\citet{C&K} grid, adopting parameters
(effective temperature/surface gravity/microturbulence/metallicity) =
(\teff/\logg/\vmicro/[A/H]) = (4900~K/2.0/2.0~\kmsec/--1.50).  
Each spectrum was corrected to the restframe system, and observed RVs were then
corrected to the heliocentric system. 
The mean RVs, together with the associated rms obtained from the
average of individual exposures, are listed in Table~1.
Since NGC\,3201 has a distinct high RV, we can safely assume all the
observed 18 stars are cluster members.
The final mean heliocentric RV for our NGC\,3201 giants is 
$\langle$RV$\rangle$=$+495.3\pm0.9$~\kmsec\ ($\sigma$=3.7~\kmsec),
which lies within 1~$\sigma$ of the literature value of
$\langle$RV$\rangle$=$+494.0\pm0.2$~\kmsec\ listed in the Harris
catalogue \citep{Harris10}.

Finally, the individual exposures for each star have been co-added.
The typical signal-to-noise ratio for the combined spectra 
around the [O\,{\sc i}] $\lambda$6300~\AA\ line ranges from S/N$\sim$120 to
$\sim$170, depending on the brightness of the star and the number of exposures.
The list of observed giants in NGC\,3201 is reported in
Table~1, together with coordinates, photometric
information, radial velocities (RVs), and number of exposures.

\section{Model atmospheres}\label{sec:atm}

The relatively high resolution and the large spectral coverage of our
spectra allowed a fully-spectroscopic estimate of the
stellar parameters, \teff, \logg, [A/H] and \vmicro.
Hence, we determined \teff\ by imposing the excitation potential (E.P.)
equilibrium of the Fe\,{\sc i} lines and gravity with the ionization
equilibrium between Fe\,{\sc i} and Fe\,{\sc ii} lines. Note that for
\logg\ we impose Fe\,{\sc ii} abundances that are
0.05-0.07~dex higher than the Fe\,{\sc i} ones to adjust for non-local
thermodynamic equilibrium (non-LTE) effects \citep{Bergemann12,
  Lind12}. This difference is justified by the non-LTE corrections to
Fe~{\sc i} and Fe~{\sc ii} derived for stars \#69 and \#82 (see
Section~\ref{sec:iron}). The final Fe abundances are based 
on Fe~{\sc i}, foor which we have more available lines.
For this analysis, \vmicro\ was set to
minimize any dependence of Fe\,{\sc i} abundances as a function of
equivalent width (EW). 

As an independent test of our results, we also derived atmospheric 
parameters from our $HST$ photometry (see
Section~\ref{sec:phot_data}).
For that purpose, the $m_{\mathrm{F438W}}$ and $m_{\mathrm{F606W}}$ magnitudes have been
converted to $B$ and $V$ \citep{JA08}, 
which we then used to estimate temperatures
from the \citet{Alonso99} color-temperature calibrations,
assuming a mean E$(B-V)$=0.24, and a mean [A/H]=$-$1.54~dex.
Surface gravities were then obtained from the apparent $V$ magnitudes,
the photometric \teff, bolometric corrections from \citet{Alonso99},
apparent distance modulus of $(m-M)_{V}=$14.20 \citep{Harris10}, and
a stellar mass of 0.70~$M_{\odot}$. 
Once \teff\ and \logg\ have been fixed from photometry, we derived
\vmicro\ from the Fe\,{\sc i} lines as explained above.

\begin{figure*}
   \centering
   \includegraphics[width=0.7\textwidth]{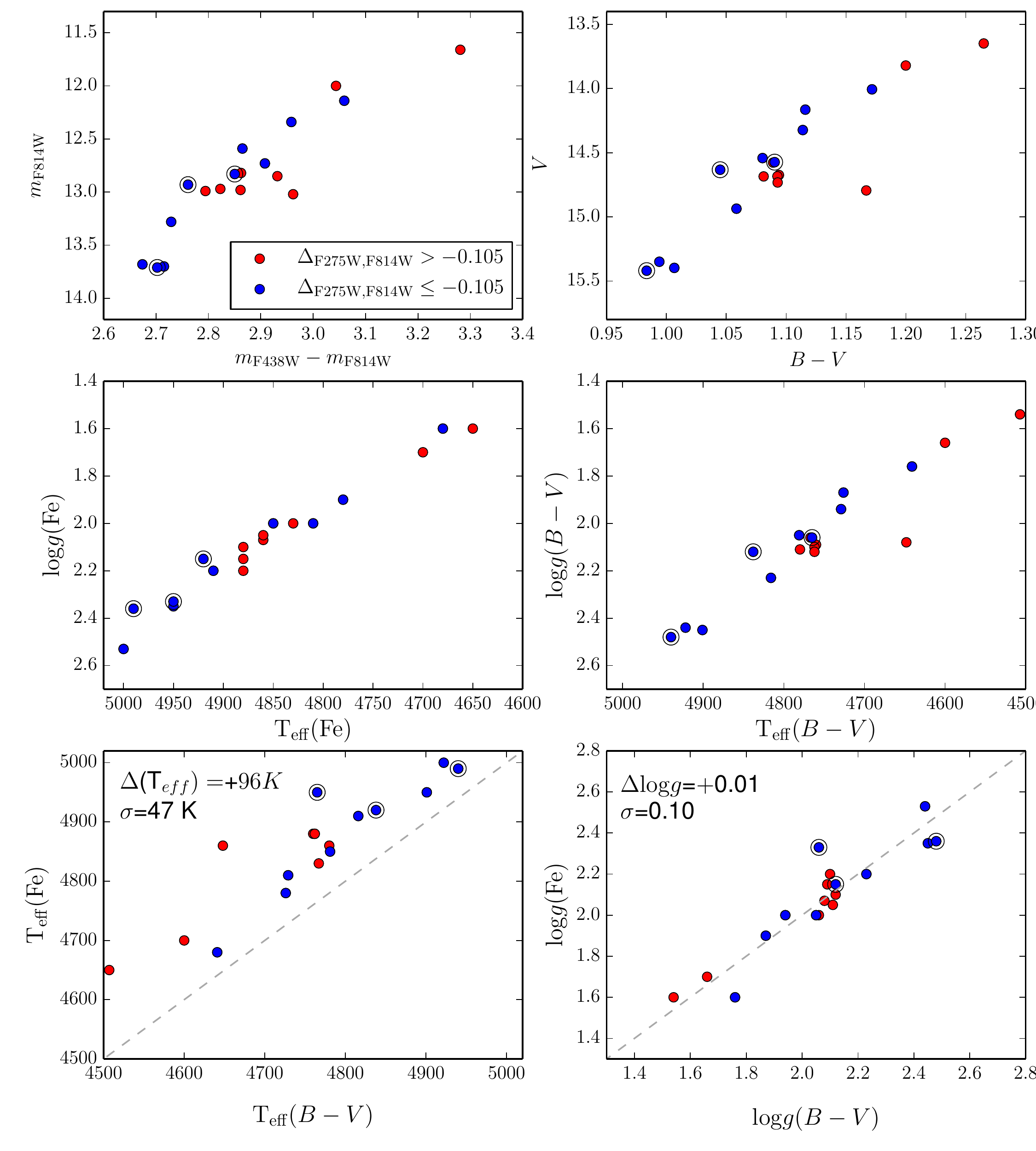}
      \caption{{\it Upper panels}: CMDs from $HST$ photometry. On the
        left we display the $m_{\mathrm {F814W}}$ vs.\,$m_{\mathrm {F438W}} - m_{\mathrm
          {F814W}}$ CMD, while on the right-hand panel 
        we show the $V$-$(B-V)$ CMD with mag obtained from
        transforming the original $HST$ mags (details in
        Section~\ref{sec:phot_data}). 
        {\it Middle panels}: \logg\ vs.\, \teff\, as obtained from the
        Fe lines (left), and from the $(B-V)$ color (right). 
        {\it Bottom panels}: Comparison between spectroscopically and
        photometrically-derived parameters, \teff\ and \logg. The dashed line represents perfect
        agreement. We label the mean difference and the rms among the two sets
        of parameters.
        In all the panels blue and red dots represent
        stars with \x$\leq$0.105 and \x$> -$0.105, respectively.}
        \label{fig:parameters}
   \end{figure*}

The top panels of Figure~\ref{fig:parameters} show the spectroscopic
targets on the $HST$ CMD (left) and the $V$-$(B-V)$ one (right), with stars
colored blue and red for \x$\leq -$0.105 and \x$> -$0.105, respectively.
The small colour offset suggests that, at a given luminosity,
the blue stars might be slightly hotter. Our adopted \logg-\teff\ values, that are
completely independent from photometry do not show significant evidence
for such a trend (left-middle panel), which might be slightly more
visible in the photometric \logg-\teff\ plane (right-middle panel).
Our internal errors in \teff\ is probably larger than the expected
difference between bluer and redder stars.

To have an estimate of the internal errors associated with our adopted
parameters 
we compare   
the \teff/\logg\ from Fe lines with those derived from
the $(B-V)$ colours in the bottom panels of
Figure~\ref{fig:parameters}. We obtain: $\Delta$\teff=\teff$_{\rm 
  {Fe~lines}}-$\teff$_{(B-V)}=+$96~K (rms=47~K), and
$\Delta$\logg=\logg$_{\rm {Fe~lines}}-$\logg$_{(B-V)}=+$0.01
(rms=0.10).  
This comparison suggests that the spectroscopic \teff\ scale is
systematically higher by $\sim$100~K (see bottom-left panel), but the
internal errors are smaller, comparable with the rms of the average 
differences, i.e. about 50~K.
The spectroscopic and photometric gravity scales agree, with a rms of
0.10~dex (bottom-right panel).
In the following we adopt typical internal
uncertainties of 50~K for \teff, 0.20~dex for \logg, 0.20~\kmsec\ for \vmicro, and
0.10~dex for metallicity, but emphasise that systematic errors might
be significantly larger.
The adopted atmospheric parameters obtained from spectroscopy 
are listed in Table~2, where we also list the inferred Fe abundances
and the \teff/\logg\ values from photometry.

\section{Chemical abundances analysis}\label{sec:abundances}

In this work we infer chemical abundances for 24 elements, namely 
Li, O, Na, Mg, Al, Si, Ca, Sc ({\sc ii}), Ti ({\sc i} and {\sc ii}), V, Cr ({\sc
  i} and {\sc ii}), Mn, Fe ({\sc i} and {\sc ii}), Co, Ni, Cu, Zn, Y
({\sc ii}), Zr ({\sc ii}), Ba ({\sc ii}), La ({\sc ii}), Pr ({\sc
  ii}), Nd ({\sc ii}), Eu ({\sc ii}).
Chemical abundances were derived from a LTE analysis by using the
 MOOG code \citep{moog}, and 
the alpha-enhanced Kurucz model atmospheres of
\citet{C&K}, whose parameters have been obtained as
described in Section~\ref{sec:atm}.
A list of our analyzed spectral lines, with their associated
equivalent widths (EWs), excitation
potentials (EPs) and total oscillator strengths (log~$gf$), is provided
in Tab.~8. 

The chemical abundances for all the elements, with the exception of 
those discussed below, have been inferred from an EW-based analysis.
We now comment on some of the transitions that we used.

{\it Lithium:}
Lithium abundances could be derived for 11 out of the 18 observed giants by
spectral synthesis of the Li\,{\sc i} $\lambda$6707~\AA\ blend using
linelist from \citet{LAB09} and \citet{H&B97}. 
The abundances have been then corrected for non-LTE effects following \citet{LAB09}.
The location on the color-magnitude diagram clearly suggests that the
stars without detectable Li are brighter than the bump, suffering
indeed from strong Li depletions that occur at this
luminosity. According to the RGB bump $V$ mag provided by \citet{Nataf13} 2 of
the 11 stars with Li measurements are fainter than the bump, hence
their abundances might not be directly compared with those in {\it
  pre-bump} giants (see Table~3). 

{\it Proton-capture elements:}
In this group of elements, we have derived abundances for O, Na, Al and Mg.
Oxygen abundances were inferred from the synthesis of the forbidden
[O\,{\sc i}] line at 6300~\AA.
Telluric O$_{2}$ and H$_{2}$O spectral absorptions often affect the O
line at 6300~\AA. 
Although telluric features have been removed as detailed in Section~\ref{sec:spec_data}
even with such a subtraction procedure, we caution that residual telluric
feature contamination might be of concern for the analysis of the
6300.3 [O\,{\sc i}] line. 
We determined Na abundances from the EWs of the Na\,{\sc i} doublets at
$\sim$5680~\AA\ and $\sim$6150~\AA, 
aluminum from the synthesis of the doublet at
$\sim$6667~\AA, and magnesium abundances from the EWs of the
transitions at $\sim$5528, 5711~\AA. 
Given the weakness of the Al doublet, for four stars we could provide
only an upper limit.
Sodium abundances have been corrected for deviations from LTE \citep{Lind2011}.

{\it Manganese:}
For Mn, we have synthetised the spectral lines at around
5395, 5420, 5433, 6014, 6022~\AA, by assuming 
f(${\phantom{}}^{55}$Mn)=1.00. When available, the hyperfine splitting
data have been taken from \citet{LawlerA, LawlerB}, otherwise from the
\citet{Kur09} compendium\footnote{{\sf http://kurucz.harvard.edu/}}.   

{\it Copper:}
Abundances for Cu were inferred from synthesis of the Cu\,{\sc i}
line at around 5105~\AA. 
Both hyperfine and isotopic splitting were
included in the analysis, using the well-studied spectral line component
structure from \citet{Kur09}. Solar-system isotopic fractions
were assumed in the computations: f(${\phantom{}}^{63}$Cu)=0.69 and
f(${\phantom{}}^{65}$Cu)=0.31.  

{\it Neutron-capture elements:} 
In the group of the {\it neutron}-capture ({\it n}-capture) elements
we derived abundances for Y, Zr, Ba, La, Pr, Nd, and Eu. 
For most of these elements we performed a spectral synthesis analysis, as
hyperfine and/or isotopic splitting and/or blending features needed to
be taken into account. 

Specifically, spectral synthesis was employed for Zr
($\lambda$5112~\AA), Ba ($\lambda$5854, 6142, and 6497~\AA), La
($\lambda$4921, 5115, 5291, 5304, 6262, 6390, 6774~\AA), Pr
($\lambda$5323~\AA), and Eu ($\lambda$6645~\AA).   
Our Ba abundances were computed assuming the \citet{McW98} 
$r$-process isotopic composition and hyperfine splitting. 
In all the other cases we have assumed the Solar-system isotopic
fractions. 

The inferred chemical abundances are listed in
Tables~\ref{tab:LiToSc}--\ref{tab:CuToEu}. 
Internal uncertainties to these abundances due to the adopted model
atmospheres were estimated by varying the stellar parameters, one at a
time, by the amounts estimated in Section~\ref{sec:atm}, namely
\teff/\logg/[A/H]/\vmicro=$\pm$50\,K/$\pm$0.20\,cgs/$\pm$0.10\,dex/$\pm$0.20\,\kmsec. 
In addition to the contribution introduced by internal errors in
atmospheric parameters, we estimated the contribution due to the
limits of our spectra, e.g.\ due to the finite S/N that
affects the measurements of EWs and the spectral synthesis. 

To estimate the contribution to the internal uncertainties given by
the quality of the spectra ($\sigma_{\rm EWs/fit}$), we have compared the EWs obtained from individual
exposures of the same stars. For this comparison we used a relatively
bright star (\#58), with five exposures, and a fainter star (\#158), with
eleven exposures. We get a typical error in EWs of $\sim$1.5~m\AA,
obtained as the average rms of the EWs measurements for each line
divided by the $\sqrt{(N-1)}$, where $N$ is the number of the exposures.
For each element, the errors in chemical abundances due to the EWs have been calculated by
varying the EWs of spectral lines by the corresponding uncertainty.
For the species inferred from spectral synthesis we have varied the
continuum at the $\pm 1~\sigma$ level, and re-derived the chemical
abundances from each line. 

For the spectral lines analysed with synthesis 
we follow the approach by \citet{JEN10} and \citet{DY13emp}. For each element, we replace the
rms ($\sigma$) in Tables~\ref{tab:LiToSc}--\ref{tab:CuToEu}
by the maximum $\sigma$ value. 
Then, we derive max($\sigma$)/$\sqrt{N_{\rm lines}}$. 
For those elements whose abundance is
inferred from just one line we use the typical uncertainty introduced by
the continuum scatter as our $\sigma_{\rm EWs/fit}$. 
Typical values obtained for each element are listed in
column 8 of Table~6. The total error is obtained by quadratically
adding this random error with the uncertainties introduced by
atmospheric parameters. 

Since the EWs/continuum placement errors are random, the uncertainty 
associated to those elements with a larger number of lines is lower.
Hence, the corresponding uncertainty associated with Fe~\,{\sc i} is
negligible, while for those species inferred from one or
two weak spectral lines, the error due to the limited S/N is dominant
(e.g.\, O, Al, Zr, Pr, and Eu).

\section{The chemical composition along the first population of NGC\,3201}\label{sec:results}

   \begin{figure*}
   \centering
   \includegraphics[width=0.96\textwidth]{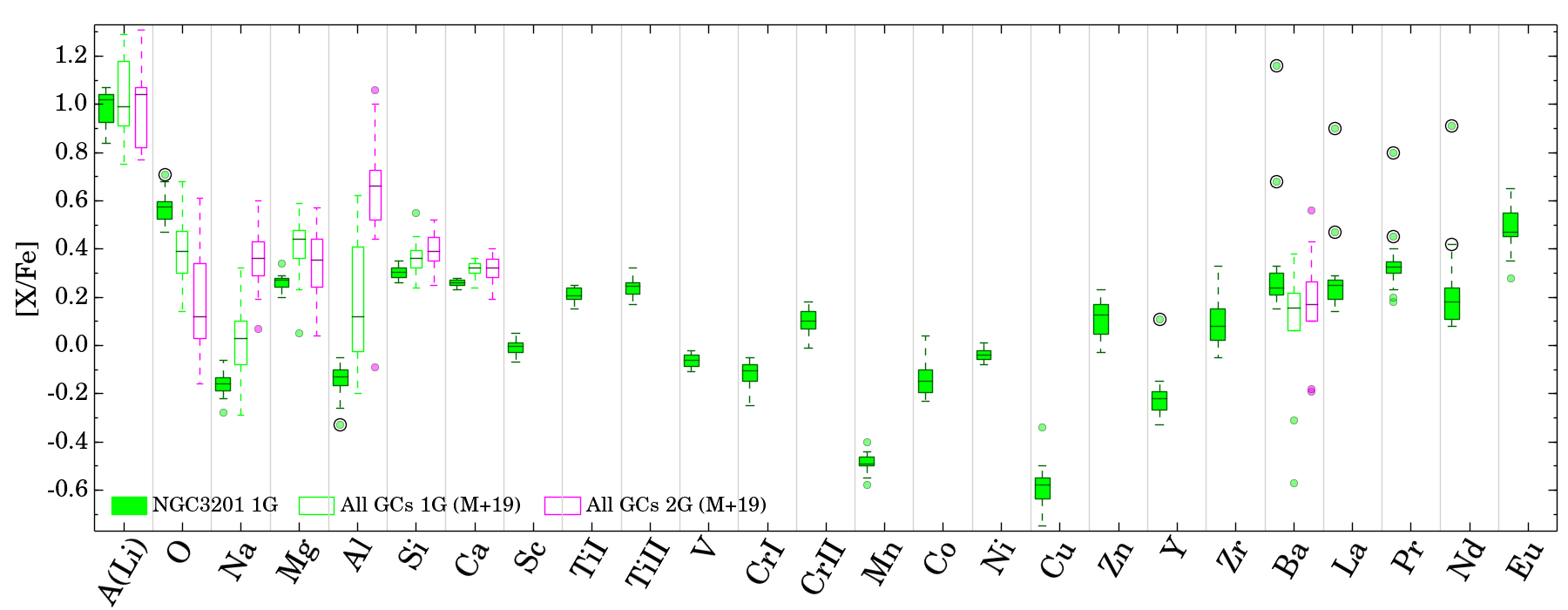}
      \caption{ Box-and-whisker plot for the chemical abundances of
        the 1G stars observed on the ChM of NGC\,3201 (green filled
        boxes). All the abundances are relative to Fe, except for Li,
        for which we use the usual A(Li) notation. Plotted Li
          abundances are corrected for non-LTE effects. 
        As a comparison, we also show the box-and-whisker plots for all the GCs for which
        abundances on ChMs are available. These
        boxes have been constructed with the average abundances for 1G
        and 2G listed in Table~2 from \citet{Mar19}.
        1G and 2G average abundances for all GCs are represented in
        green and magenta empty boxes. Each box represents the
        interquartile range (IQR) of the distribution, with the median
        abundance marked by an horizontal line. The whiskers include
        observations that fall below the first quartile minus 1.5$\times$IQR
        or above the third quartile plus 1.5$\times$IQR. 
        As the majority of the used literature studies for Na report
        LTE abundances, the plotted [Na/Fe] values for NGC\,3201 are not
        corrected for non-LTE. Small filled circles
        represent outliers in the data. Outliers highlighted with
        black open circles are the candidate binaries (see
        Section~\ref{sec:binarity} for details). } 
        \label{fig:box}
   \end{figure*}
   \begin{figure*}
   \centering
   \includegraphics[width=0.9\textwidth]{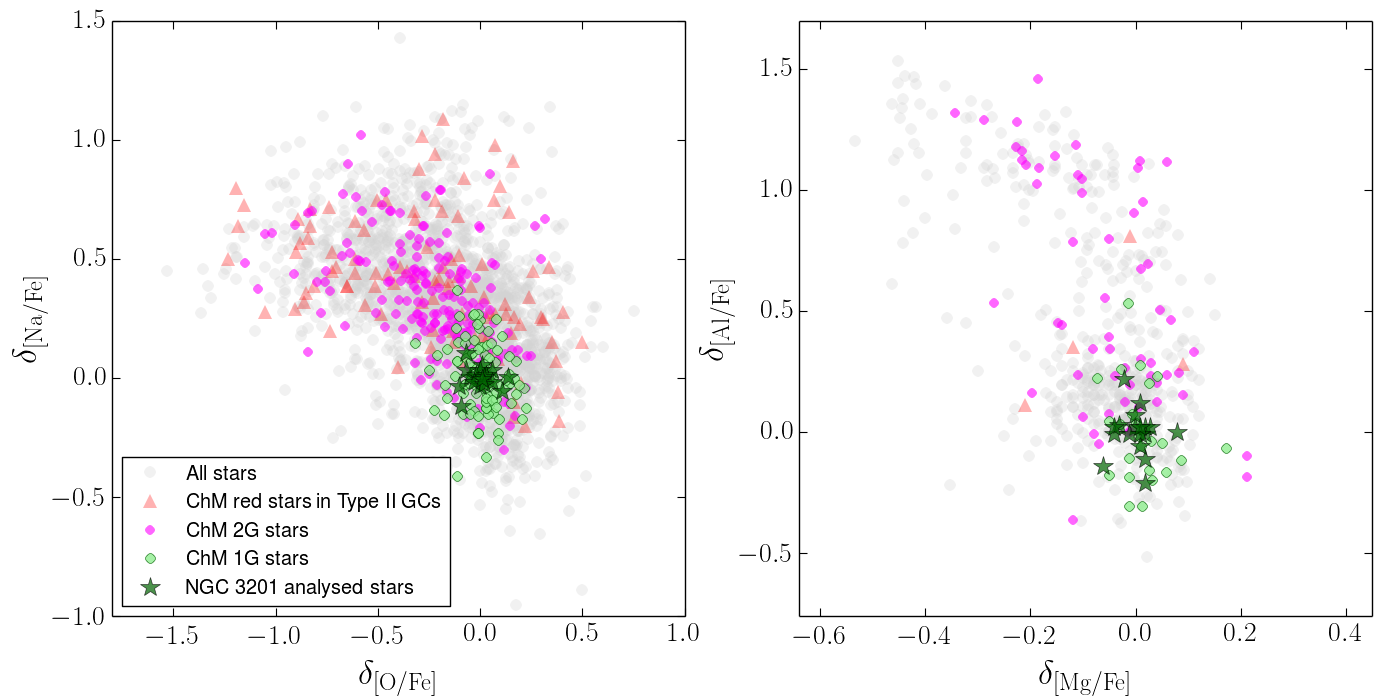}
      \caption{Abundance ratios of $\delta_{\rm [Na/Fe]}$ as a function
        of $\delta_{\rm [O/Fe]}$ (left) and $\delta_{\rm [Al/Fe]}$ as a function
        of $\delta_{\rm [Mg/Fe]}$ (right) for all the GCs analysed in
        \citet{Mar19}. 
        The plotted abundances are relative to the average
        abundances of 1G stars, as defined by Marino and
        collaborators. Different symbols and colors represent stars
        belonging to different populations on the ChM. The abundances
        of the analysed stars in NGC\,3201 perfectly overlap with the
        ChM 1G abundances on these $\delta_{\rm [O/Fe]}$-$\delta_{\rm
          [Na/Fe]}$ and $\delta_{\rm [Mg/Fe]}$-$\delta_{\rm
          [Al/Fe]}$ planes.} 
        \label{fig:nao}
   \end{figure*}

In this section we explore the chemical composition in different
elemental species along the elongated 1G observed on the ChM of
NGC\,3201. We emphasize that, although the range covered by our
targets in the \x\ axis is large, they span only a small range in \y, and
their position on the map is consistent with the 1G, as defined by
\citet{Mil15, Mil17}. As the \y\ axis of the ChM is shaped by light
elements variations, primarily driven by the N enhancements in the 2G
stars, we expect homogeneous abundances in this chemical
specie \citep{Mar19}. 

Overall, for our 18 giants we obtain a mean iron abundance of
[Fe/H]=$-$1.50$\pm$0.02~dex (rms=0.07~dex), consistent with the value
of [Fe/H]=$-$1.59 listed in \citet{Harris10}.
Figure~\ref{fig:box} shows a summary of the other chemical abundance
ratios obtained for our 1G sample of NGC\,3201 stars\footnote{Chemical
  abundances are expressed in the standard notation, as the
  logarithmic ratios with respect to solar values, [X/Y]=$\mathrm
  {log(\frac{N_{X}}{N_{Y}})_{star}-log(\frac{N_{X}}{N_{Y}})_{\odot}}$. For
lithium, abundances are reported as A(Li)=$\mathrm {log(\frac{N_{Li}}{N_{H}})_{star}}+12$.}.
Besides observing the typical chemical pattern of Population~II stars, namely
the enhancement in the $\alpha$-elements (Mg, Si, Ca, Ti), and the
typical solar-scaled abundances of Fe-peak elements, more interesting
for this study is the investigation of the elements that have a
close connection with the shape of the ChM. 

In Figure~\ref{fig:box} we show a comparison of our results for
NGC\,3201 with all the GCs for which abundances on ChMs are available
from \citet[Table~2]{Mar19}. Overall, our inferred abundances agree
with those observed in the 1G as selected on the ChM 
of GCs, as expected from the location of our targets along the ChM of NGC\,3201.
In the next sections we will discuss all the
interesting abundance patterns we observe in NGC\,3201, focusing on
those elements that play an important role in the multiple stellar
populations phenomenon and in shaping the ChM. In particular,
\citet{Mar19} have analysed the chemical abundance pattern along the
ChM, for the species most involved in the multiple stellar population
phenomenon, namely the $p$-capture (e.g. O, Na, Mg, Al) and
$n$-capture elements (Ba). 

First, we note that the abundances relative to Fe ([X/Fe])
are generally consistent with uniform chemical abundances in the
plotted elements. By comparing the observed rms associated to the
mean average abundances, as listed in Tables~3--5, with the estimated
errors (Table~6), it appears that, in most cases, our expected
uncertainties are higher. This suggests that our estimated errors
might be overestimated. An exception to this general trend are the
$n$-capture elements, that will be discussed in the following sections. 

As already mentioned, the chemical abundances of the
18 stars of NGC\,3201 analysed here are consistent with the chemical
composition of 1G stars. 
Lithium abundances have a range of $\sim$0.2~dex, from A(Li)$\sim$0.75
up to A(Li)$\sim$1.0~dex, which compares to typical values of RGB stars
having experienced the full first dredge up but not having reached yet
the RGB bump level \citep{Lind09}. As highlighted in Table~3,
stars with only upper limits plus two other stars with Li measurements
are brighter than the RGB bump.
In the context of multiple stellar populations, 
\citet{Mar19} did not find strong evidence for variations of Li
between 1G and 2G stars on the ChM, with the exception of GCs like NGC\,2808, and
$\omega$~Centauri.
Hence, the comparison of Li with the average abundances of 1G and 2G
stars in all the GCs, as shown in Figure~\ref{fig:box} 
does not likely provide clear information on the population {\it
  status} of our sample in NGC\,3201. 

Oxygen and sodium are the best tracers of the multiple stellar
populations phenomenon and their abundances are indicative of
different locations along the ChM.
Specifically, for our NGC\,3201 giants
the oxygen abundances relative to iron are super-solar, as typical of
Population~II stars, while sodium is not enhanced. Both these elements
show distributions compatible with the 1G ones.
The association of all the analysed stars with the chemical composition of
the ChM 1G is clear by looking at their location on the $\delta_{\rm
  [O/Fe]}$-$\delta_{\rm [Na/Fe]}$ plane plotted in
Figure~\ref{fig:nao}, where the $\delta$ abundances are the relative
abundances to the average chemical content of 1G stars in each cluster
\citep[see][for a detailed discussion]{Mar19}. 

Chemical enhancements in aluminum, coupled with depletions in magnesium are observed in a few
clusters, with NGC\,2808 being a noticeable example. Both our Mg and
Al abundances fall in the range spanned by 1G stars.
Magnesium relative to Fe is super-solar, as expected from
a typical $\alpha$-element, and the Al abundaces distribute on the
lower abundance tail spanned by 1G stars.
Both these element distributions do not show any evidence for internal
variations within our 18 analysed stars.
The location of our stars on the $\delta_{\rm [Mg/Fe]}$-$\delta_{\rm
  [Al/Fe]}$ plane, constructed by using relative abundances similar to
those used for $\delta_{\rm [O/Fe]}$ and $\delta_{\rm [Na/Fe]}$, are
plotted in the right panel of Figure~\ref{fig:nao}. 
 
\section{Binarity and Blue stragglers}\label{sec:binarity}

   \begin{figure*}
   \centering
   \includegraphics[width=0.98\textwidth]{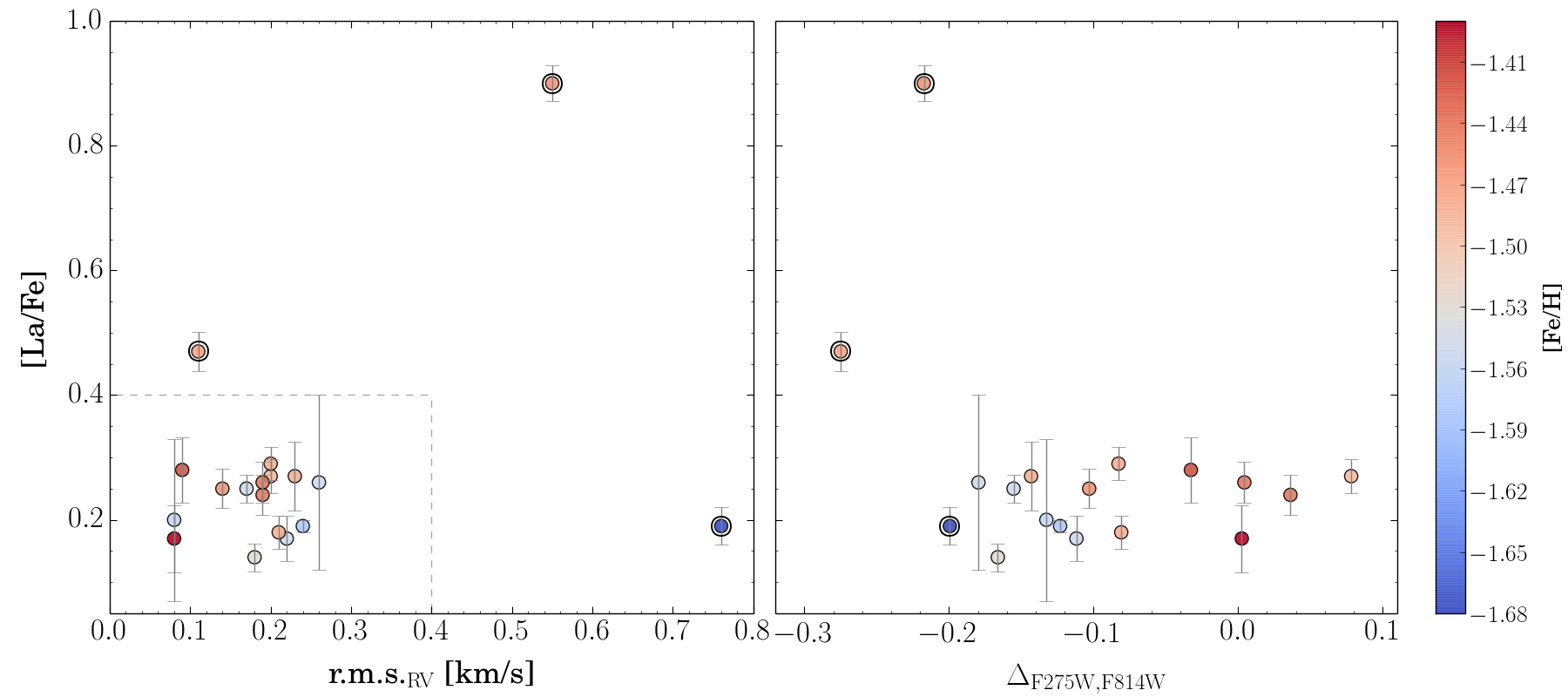}
      \caption{The abundance of the $n$-capture element La relative to
      Fe, as a function of the r.m.s. in the RVs (left panel), and
      \x\ (right panel). Error bars are the statistical
      errors associated to the average abundances for each star
      obtained from different lines. 
      The color of each star is indicative of its
      inferred Fe, as illustrated in the right-side colour bar. Our
      three binary
      candidates, outlined with black circles, lie outside the box
      delimited by a dashed line in the left panel. }
        \label{fig:LaVsRV}
   \end{figure*}

   \begin{figure*}
   \centering
   \includegraphics[width=0.98\textwidth]{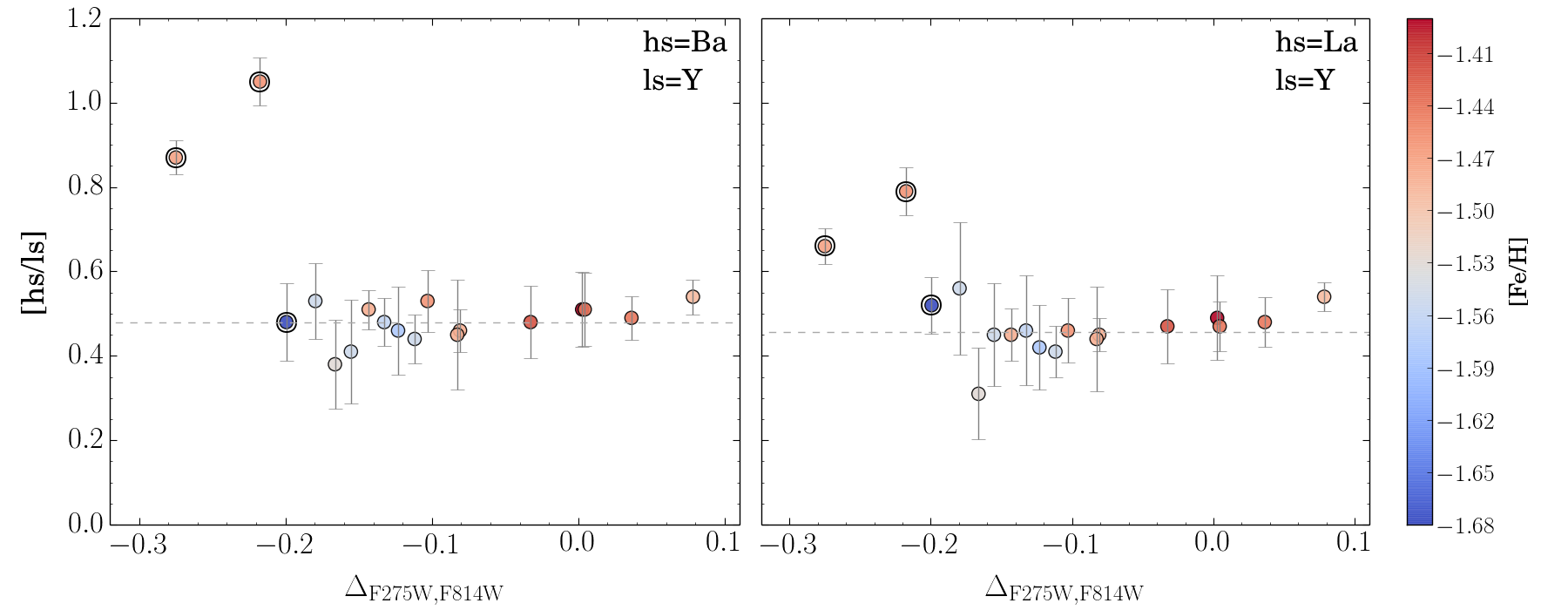}
      \caption{The abundance of [Ba/Y] (left) and [La/Y] (right),
        representative of the heavy to light $s$-element abundances
        ([hs/ls]), as a function of \x. 
        The dashed lines in both panels highlight the average
        abundance ratios for the entire sample, neglecting the three (encircled)
        binary candidate. As in Figure~\ref{fig:LaVsRV}, the color of
        each star is indicative of its inferred Fe, as illustrated in
        the right-side colour bar. }
        \label{fig:hsls}
   \end{figure*}

Noticeably, Figure~\ref{fig:box} shows a few stars with significantly
higher abundances in the $n$-capture elements, Y, Ba, La, Pr and Nd. 
This class of chemical species plays an important role in the Type~II
GCs, those displaying additional ChMs on the red side of the main
map. Stars on the red side of ChMs are typically enhanced in both Fe
and the elements mostly produced via $s$-process reactions, e.g. Ba
and La, but not in Eu, a typical $r$-process element.
Nevertheless, NGC\,3201 does not belong to the class of Type~II GCs, as it
does not show any clear additional redder sequence in the ChM as M\,22, M\,2, or
$\omega$~Centauri do \citep{Mil17}.

To investigate the nature of the variations in $s$-process elements in
NGC\,3201, we plot in Figure~\ref{fig:LaVsRV} the abundance of [La/Fe]
taken as representative of the $s$-process elements, as a function of
the r.m.s. in the RVs (left-hand panel) and \x\ (right-hand panel).  
While most stars show typical [La/Fe] abundances around 0.25~dex, two 
stars have [La/Fe]$>$0.40~dex, with one reaching an abundance as high
as [La/Fe]$=$0.90~dex. Interestingly, this extremely-La rich star
(\#93) also has a high r.m.s. in RV ($\sigma_{\mathrm {RV}}$),
compared to the bulk of the stars lying around 0.2~\kmsec. 
A third outlier in the [La/Fe]-$\sigma_{\mathrm {RV}}$ plane displays
the highest r.m.s. in RV (0.76~\kmsec, star \#67), but it is not an outlier
for [La/Fe]. 

The high RV r.m.s. in two giants of our sample suggests that
these stars are indeed binaries. 
By applying a maximum likelihood approach, similar to that used in
\citet{Frank15} and \citet{Piatti}, assuming a typical uncertainty of
0.17~\kmsec, we found that the intrinsic dispersions in the RVs of the
stars \#93 and \#67 are 0.51 and 0.74~\kmsec, respectively. Such
values, higher than zero, corroborate the idea that these stars are
binary candidates. 
The very high abundance in $s$-elements of one of these stars is
consistent with the binary nature of this object. Similarly, star
\#149 having relatively high abundances in $s$-elements can be
regarded as a binary candidate, although our data do not show any
evidence of high variations in its RV. 

As shown in Figure~\ref{fig:hsls}, the stars with high $n$-capture
element abundances also show higher [hs/ls] than the mean value of
$\sim$0.5~dex, suggesting that they are likely the result of direct
mass transfer with a low-mass asymptotic giant branch (AGB)
companion. The abundance ratios [La/Y] and [Ba/Y], as a function of
\x, are indicative of the heavy to light $n$-capture element abundance
ratio, which is sensitive to the neutron exposure and neutron
density. Higher [hs/ls] is indicative of the operation of the
$\phantom{}^{13}$C($\alpha$,$n$)$\phantom{}^{16}$O neutron source
operating in low-mass AGB stars \citep[e.g.,][]{Gallino98,
  Fishlock14}, where values around zero or negative values indicate
the $\phantom{}^{22}$Ne($\alpha$,$n$)$\phantom{}^{25}$Mg reaction. The
higher [hs/ls] of the two giants enriched in $n$-capture elements
might suggest direct mass transfer between the star we now observe and
previous low-mass, low-metallicity AGB stars
\citep[e.g.\,][]{Cristallo09, K&L14}.   

In the last column of Table~1 we list the maximum time interval between
the observations of each star. We note that some stars, including
\#149, have been observed over a shorter time interval (3 days rather
than months), which might have prevented us from detecting RV
measurable variations in longer-period binaries. 

In the following we will consider these three stars as binary
candidates. Their locations in
Figures~\ref{fig:parameters},~\ref{fig:box},~\ref{fig:LaVsRV} (as well
as in some of the following figures) are
highlighted with black open circles.
We warn the reader that we cannot rule out the presence of 
other binaries with no obvious $s$-elements or/and RV differences
among our remaining 15 giants.  

\begin{figure*}
   \centering
   \includegraphics[width=0.6\textwidth]{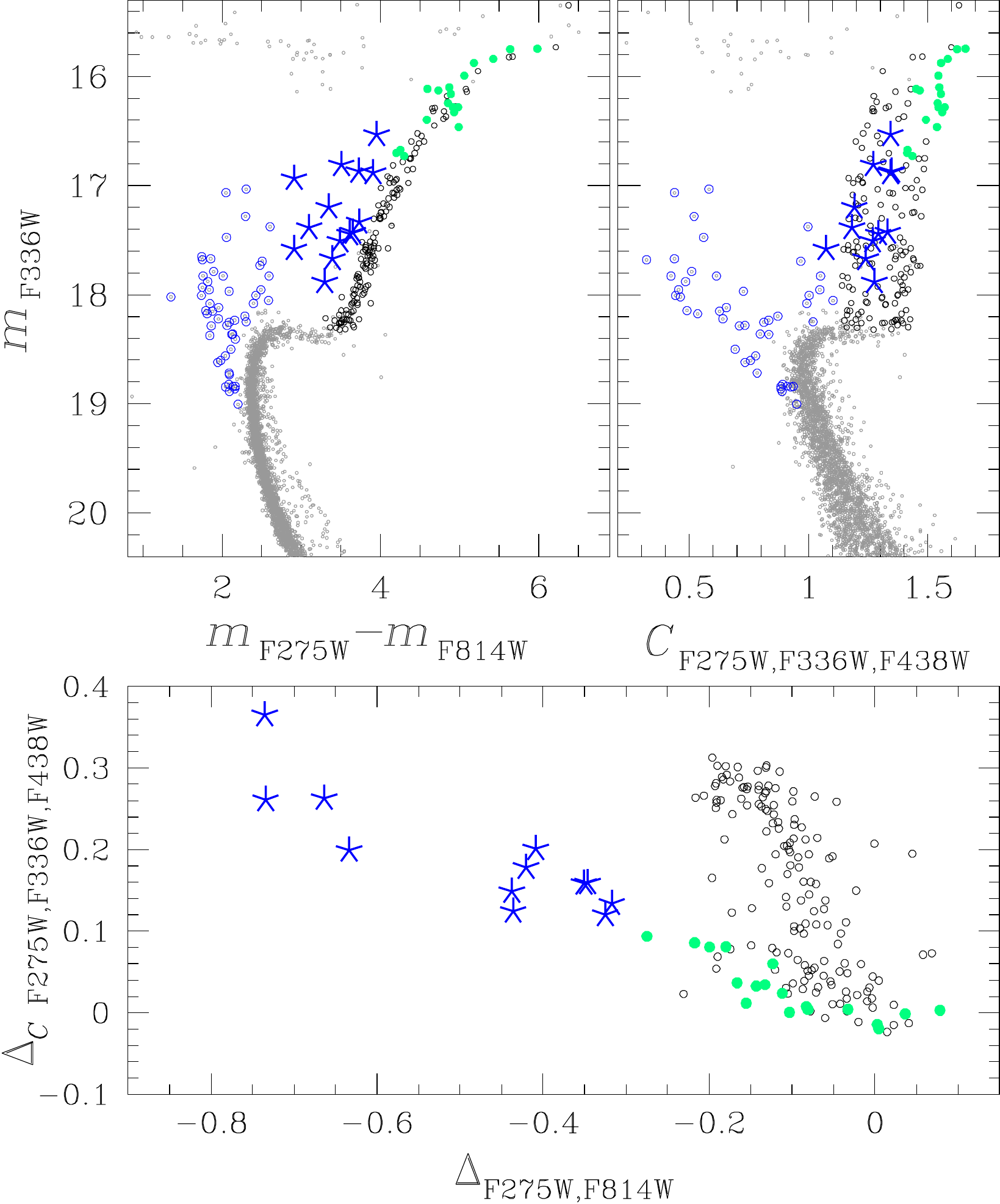}
      \caption{{\it Upper panels}: $m_{\mathrm {F336W}}$-$(m_{\mathrm
          {F275W}}-m_{\mathrm {F814W}})$ CMD (left) and $m_{\mathrm {F336W}}$-$C_{\mathrm
          {F275W,F336W,F438W}}$ diagram (right panel) of
        NGC\,3201. {\it Lower panel}: ChM of NGC\,3201, similar to
        Figure \~1, but extended to show the evolved BSSs. In all the 
        panels the green dots represent our analysed spectroscopic
        targets, the open blue circles, the blue stragglers, and the
        star-like symbols are the evolved blue stragglers.}
        \label{fig:CMD_BIN}
   \end{figure*}

Another stellar population that may affect the appearance of the chromosome maps, and
likely has a connection to binary stars, are the blue stragglers (BSs). In
particular, we consider their evolved counterparts which lie close to the
red giant branch. In Figure~\ref{fig:CMD_BIN} we use blue symbols to
represent the candidate BS stars of NGC\,3201 in the $m_{\mathrm
  {F336W}}$-$(m_{\mathrm {F275W}}-m_{\mathrm {F814W}})$ CMD. 
The objects marked with blue starred symbols are possible BSs,
selected on the CMD, that are
evolving towards the RGB phase, which clearly display lower
$C_{\mathrm {F275W,F336W,F438W}}$ values 
in the $m_{\mathrm {F336W}}$-$C_{\mathrm {F275W,F336W,F438W}}$ diagram
(upper-right panel). 
Blue stragglers in GCs are tightly linked to the cluster
binary populations 
\citep{2009Natur.457..288K}. Blue stragglers 
in open clusters themselves have a very high binary fraction
\citep{2009Natur.462.1032M}, and preliminary evidence suggests the same
is true in NGC\,3201 \citep{Giesers19}. When we
include the evolved BSs in the ChM (lower panel of
Figure~\ref{fig:CMD_BIN}), they populate a well-defined sequence on
the blue extension of the 1G sequence.  
We conclude that, based on RVs and  
$s$-process element abundance, the three bluest 1G stars that we
analyzed spectroscopically are binary systems. Some of them are
possibly associated with the BS population of NGC\,3201, as
evolved BSs would contribute to the bluest extension of the 1G
sequence in the ChM.  

\subsection{Non-interacting binaries simulations}\label{sec:simu}

\begin{figure*}
   \centering
   \includegraphics[width=0.7\textwidth]{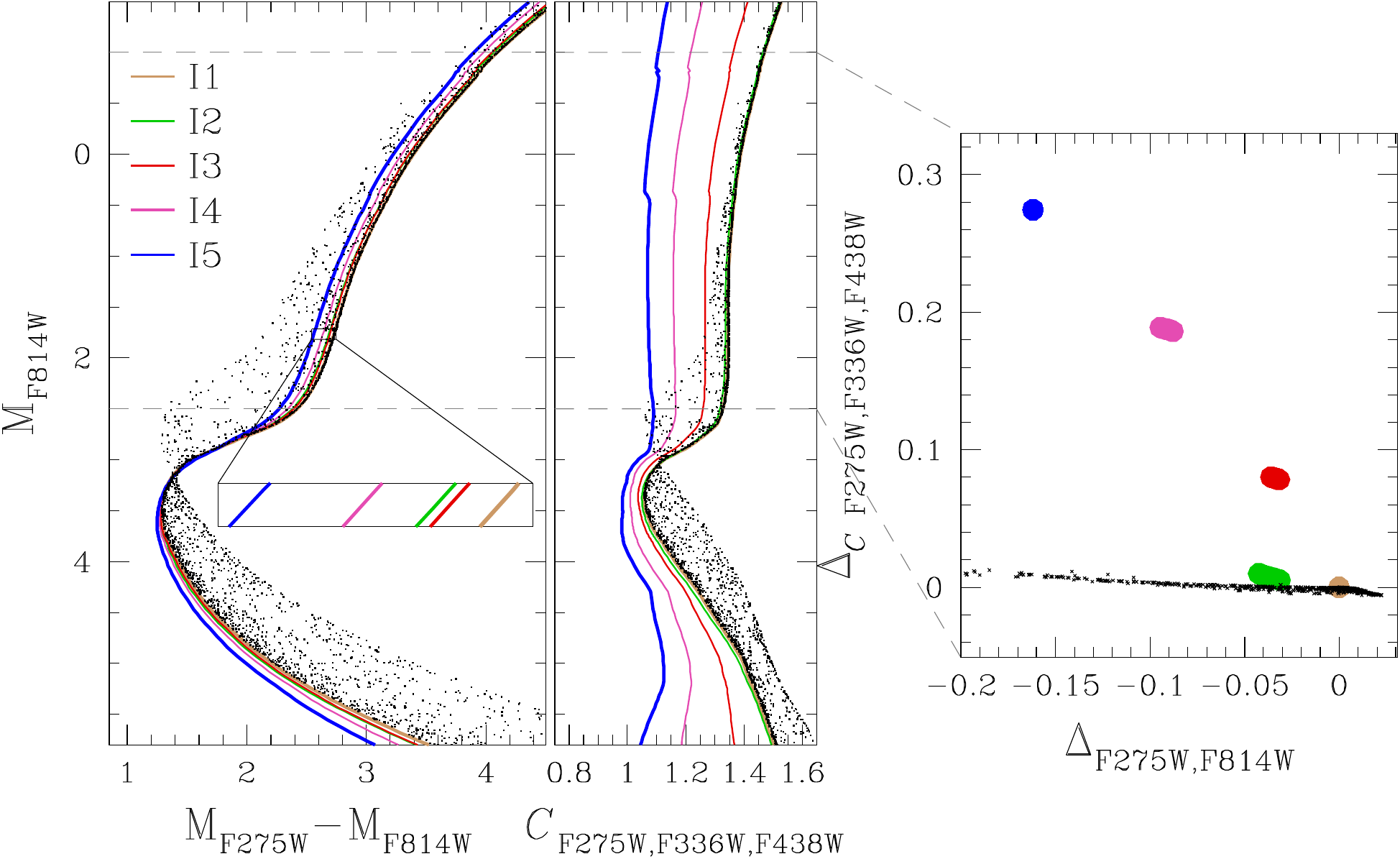}
      \caption{{\it Left panels}: Simulated CMD, with five isochrones corresponding to
        different stellar populations with different He, C, N, O:
I1 is the isochrone corresponding to the first population;
I2 has the same abundance of C, N and O as I1 but different He;
in I3, I4, and I5 abundances of He, C, N, O have been varied
to represent the second population chemical pattern.
{\it Right panel}: Location of the different stellar simulated five pupulations
at a reference magnitude on the ChM plane. 
The black points are non interacting binaries formed by pairs of I1-I1 stars.
Binaries can contribute to the width of the 1G in the ChM, but only
binaries with large mass ratio (q$\gtrsim$0.8) can provide a significant color spread. 
}
        \label{fig:CMD_SIMU}
   \end{figure*}

To investigate the effect of non-interacting binaries formed by pairs
of 1G stars, we reproduce in Figure~\ref{fig:CMD_SIMU} five
isochrones, I1---I5, with [Fe/H]=$-$1.50, [$\alpha$/Fe]=0.4 and age of
13~Gyr from \citet{Mil18}. 
These isochrones are constructed with a different combination of He, C, N, and
O, to represent the typical chemical pattern of different stellar
populations in GCs.
We used the isochrone, I1, which correponds to the 1G, to generate a
stellar population of 1G-1G binaries that we represent with black
points in Figure~\ref{fig:CMD_SIMU}. 

The corresponding ChM plane of RGB stars for the five populations, is
shown on the right panel.
Togheter with the five isochrones with different chemical abundances,
we plot a population of simulated 1G-1G binaries (black points)
constructed by assuming a flat distribution in the mass ratio (q). 
This test clearly demonstrates that binaries can contribute to the
width of the 1G in the ChM. However, we find that the shift towards
bluer \x\ is dependent on the mass ratio and only  
binaries with large mass ratio (q$\gtrsim$0.8) are able to provide a
significant color spread.

\begin{figure}
   \centering
   \includegraphics[width=0.45\textwidth]{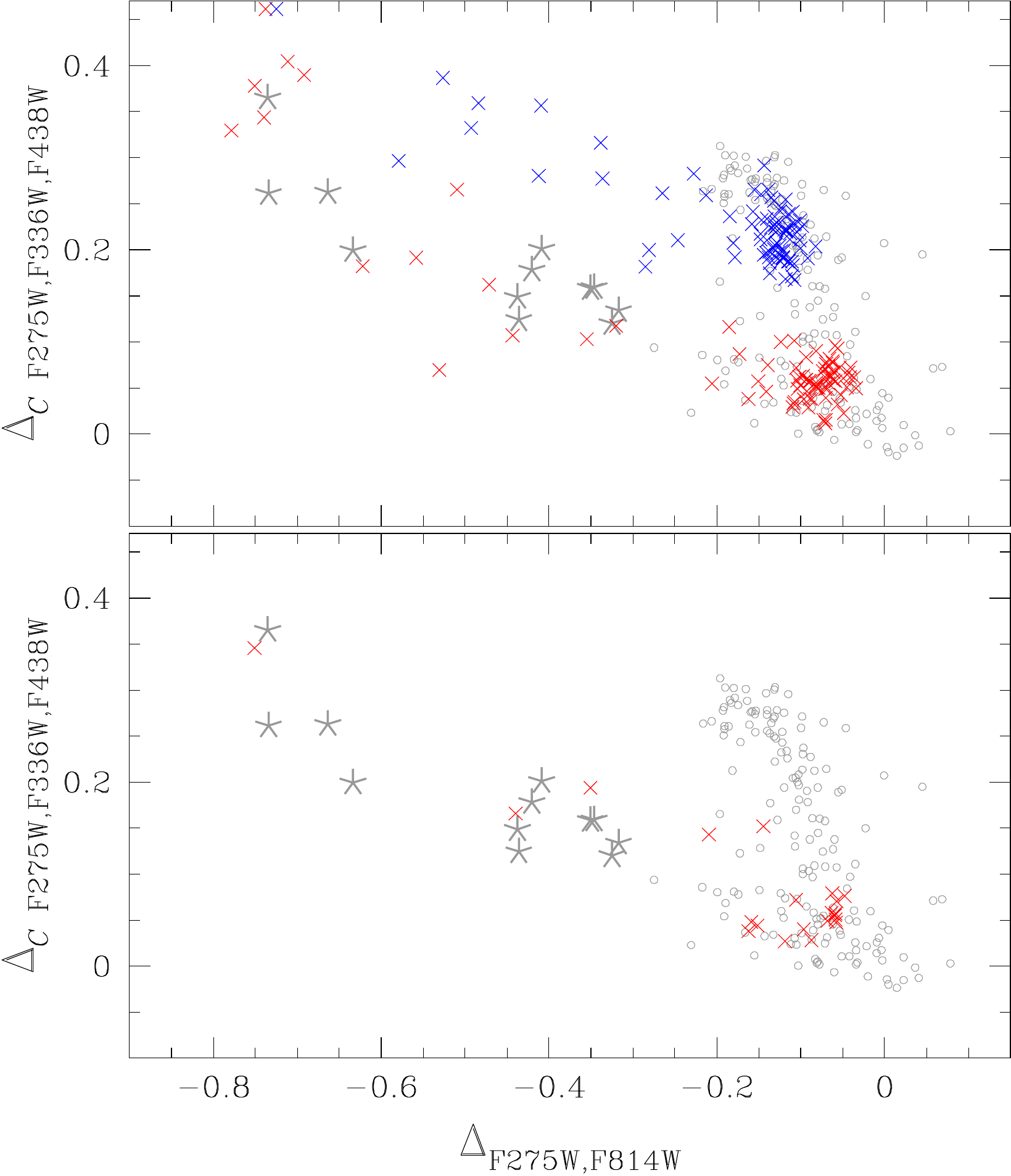}
      \caption{Observed ChM of NGC\,3201 (grey symbols). The stars
        with extremely low \x\ are plotted as grey star-like
        symbols. Superimposed to the observed ChM are simulations of
        binary stars: in the upper panel we have simulated a cluster
        with the 100\% of binaries, both the 1G-1G (red) and the
      2G-2G (blue) binaries; in the lower panel, we have simulated the 12.8\% of
      binaries, which is the observed fraction on the main sequence
      \citep{APM12bin}, by assuming all of them being 1G-1G binaries.} 
        \label{fig:binaries}
   \end{figure}

To have a more direct comparison with the ChM of NGC\,3201, in 
Figure~\ref{fig:binaries} we superimpose to the observed map two
simulated binary populations: {\it (i)} a population of binaries that
account for the whole cluster (100\% of binaries), with both the 1G-1G
and the 2G-2G pairs (upper panel); {\it (ii)} a more realistic
simulation, where the binary population fraction is 12.8\%, as
reported in \citet{APM12bin} for the main sequence binaries (lower
panel). In the latter we assume that all the binaries are 1G-1G pairs.
Again we find that the 1G-1G binaries can affect the elongation of the
1G on lower \x\ values, but the observed binary fraction on
the main sequence are not able to account for the relatively high
number of stars observed on the bluer extention of the ChM.

\section{Chromosome map and abundance pattern of 1G stars}\label{sec:iron}

On the right-hand panel of Figure~\ref{fig:LaVsRV}, we show
the [La/Fe] abundances as a function of the \x\ axis of the ChM. 
The three binary candidates are the stars with
the lowest \x\ values in our sample. 
As discussed in the previous section, this suggests that the stars
with the lowest \x\ on the map are likely associated 
with binaries and/or BSs, with no need to invoke chemical abundances variations
like He enhancements.

However, our simulations suggest that we would need a high number of
binaries at high q to account for all the stars in the 1G population.
This requirement might suggest that some additional mechanism might be
necesary to account for a range of $\sim$0.2~mag in \x\ among 1G
stars in NGC\,3201, and similarly in many other GCs. 

In Figure~\ref{fig:LaVsRV} the used color code is indicative of the Fe
abundance. We immediately note that the stars with lower \x\ also have
lower iron, with \#27 (\x$=+$0.0026) and \#33 (\x$=-$0.1231) being the
stars with the highest and lowest metallicity, respectively, giving a
maximum range of 0.19~dex (neglecting the three binary candidates).  

This result suggests that there is a small Fe variation among 1G stars
in NGC\,3201. By dividing our sample of 15 stars on the basis of the
\x\ value, we find that eight stars with \x$< -$0.10 have
$<$[Fe/H]$>$=$-$1.53$\pm$0.02, while the seven stars at higher \x\
have $<$[Fe/H]$>$=$-$1.45$\pm$0.01, more than 2.5$\sigma$ difference.
This difference is smaller than that inferred for the blue- and
red-RGB stars in the Type~II GCs \citep{Mar19}, and hardly detectable.
Our relatively high-S/N and high-resolution UVES spectra for 15 stars
allow us to detect the difference still at a $<$3~$\sigma$ level.
We also note that the dispersion associated with of our [Fe/H] average
abundance is comparable with the estimated error.

It is worth, at this point, to discuss the impact of non-LTE
corrections on this result, which is based on LTE abundance
analysis. We first note that, as amply discussed in \citet{Kovalev19},
non-LTE analysis changes the mean abundance ratios in clusters, but
intra-cluster abundance dispersions should not be significanltly
affected. However, to test this we derived the non-LTE corrections to
our Fe~{\sc i} and Fe~{\sc ii} spectral lines \citep{Bergemann12,
  Lind12}, by using the online available database ({\sf
  http://inspect-stars.com/}). For this test we use stars \#69 and
\#82, that have similar atmospheric parameters (both from spectroscopy
and photometry), but different \x, with star \#82 having a higher
derived [Fe/H]. We find that the non-LTE correction to apply to
Fe~{\sc i} lines is $+$0.07~dex for both stars, and does not change
significantly from line to line. Fe~{\sc ii} abundances decrease by
$-$0.01, again in both stars. Using non-LTE Fe abundances for the
determination of the atmospheric parameters only marginally changes
the spectroscopic temperature, by $+$30-40~K in both stars. From this
test, it is clear that, although non-LTE abundances are more
realistic, in our sample of stars the corrections are systematic and
similar in all the stars. The difference in [Fe/H] between the two
stars is kept unchanged.     

\begin{figure}
   \centering
   \includegraphics[width=0.47\textwidth]{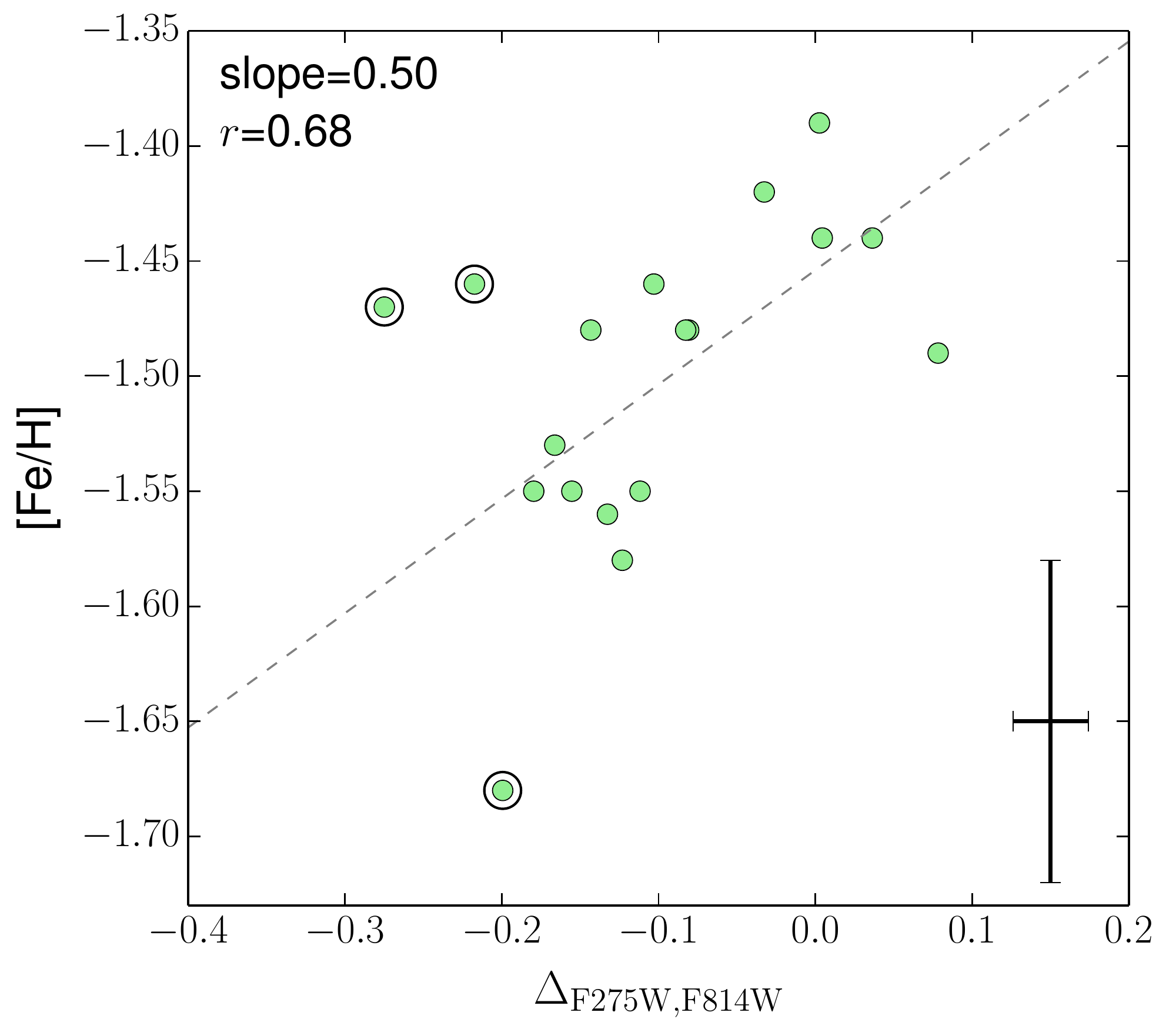}
      \caption{Chemical abundances [Fe/H] as a function of
        \x\ values on the ChM of NGC\,3201. The dashed line is the
        least squares linear fit to the data (neglecting the three
        binary candidates). The slope and the Spearman correlation
        coefficient ($r$) are highlighted in the top-left corner.} 
        \label{fig:FeX}
   \end{figure}

Figure~\ref{fig:FeX} shows [Fe/H] as a function of \x. Excluding
the three binary candidates we get a positive correlation (slope=0.50), with a
Spearman correlation coefficient $r$=0.68.
We have derived the significance of the correlation from a
Monte Carlo of 1000 realizations of our dataset composed
of 15 stars. In each realization we have assumed the observed \x\ and
a uniform Fe abundance with the associated error 
estimates of Table~6, and derived the slope. We have calculated
the fraction ($f$) of realizations where the slope is equal or higher than the
observed one, and assumed this value as the probability that the slope
is due to randomness. For [Fe/H] we get a slope higher than the
observed value in  2.2\% of the simulations (see Table~7).

\begin{figure}
   \centering
   \includegraphics[width=0.48\textwidth]{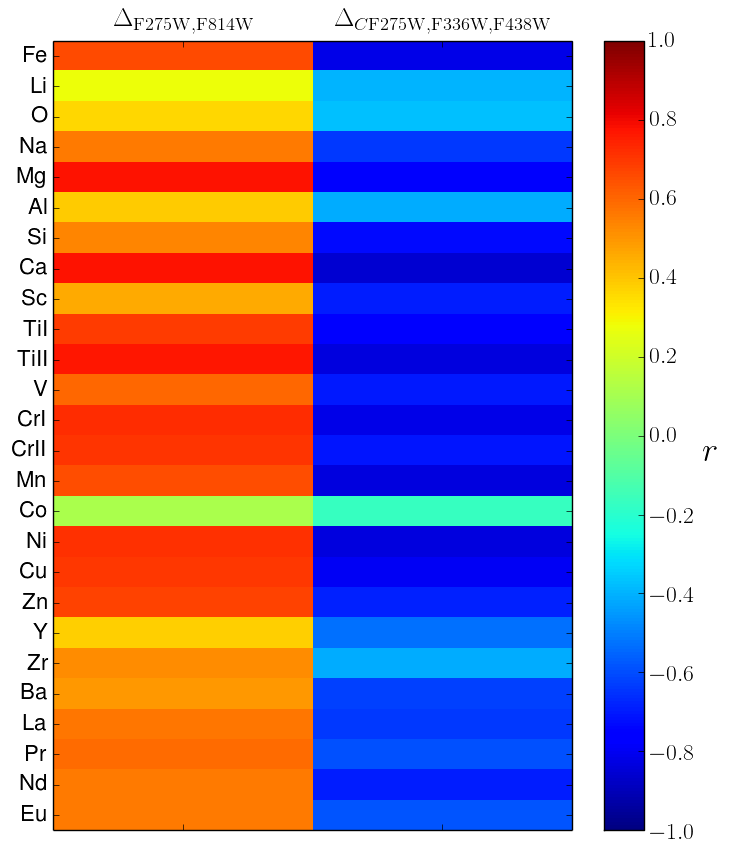}
      \caption{Chemical abundances log$\epsilon$(X) as a function of
        \x\ and \y\ values on the ChM of NGC\,3201. 
        For Li and Na we use the non-LTE abundances.
        The color code is indicative of the Spearman correlation
        coefficient ($r$). The significance of these correlations is
        listed in Table~7.}  
        \label{fig:corr}
   \end{figure}

Figure~\ref{fig:corr} represents the Spearman correlation for each
log$\epsilon$(X) abundance as a function of the \x\ and \y\ value on
the ChM of NGC\,3201. The obtained correlation values between \x\ and
the log$\epsilon$ abundances are listed in Table~7, 
togheter with the associated $f$. 
Although not all the correlations are significant, given the
relatively small sample size, clearly all the 
abundances appear to be positively correlated with \x. We regard the
negative correlation with \y\ as simply due to the fact that our stars
with low \x\ have slightly higher \y\ (see Figure~\ref{fig:targets}).
However, the presence of a general correlation between each absolute
abundance and the \x\ corroborates the presence of a small 
spread in the overall metallicity among 1G stars in NGC\,3201.

\begin{figure*}
   \centering
   \includegraphics[width=0.9\textwidth]{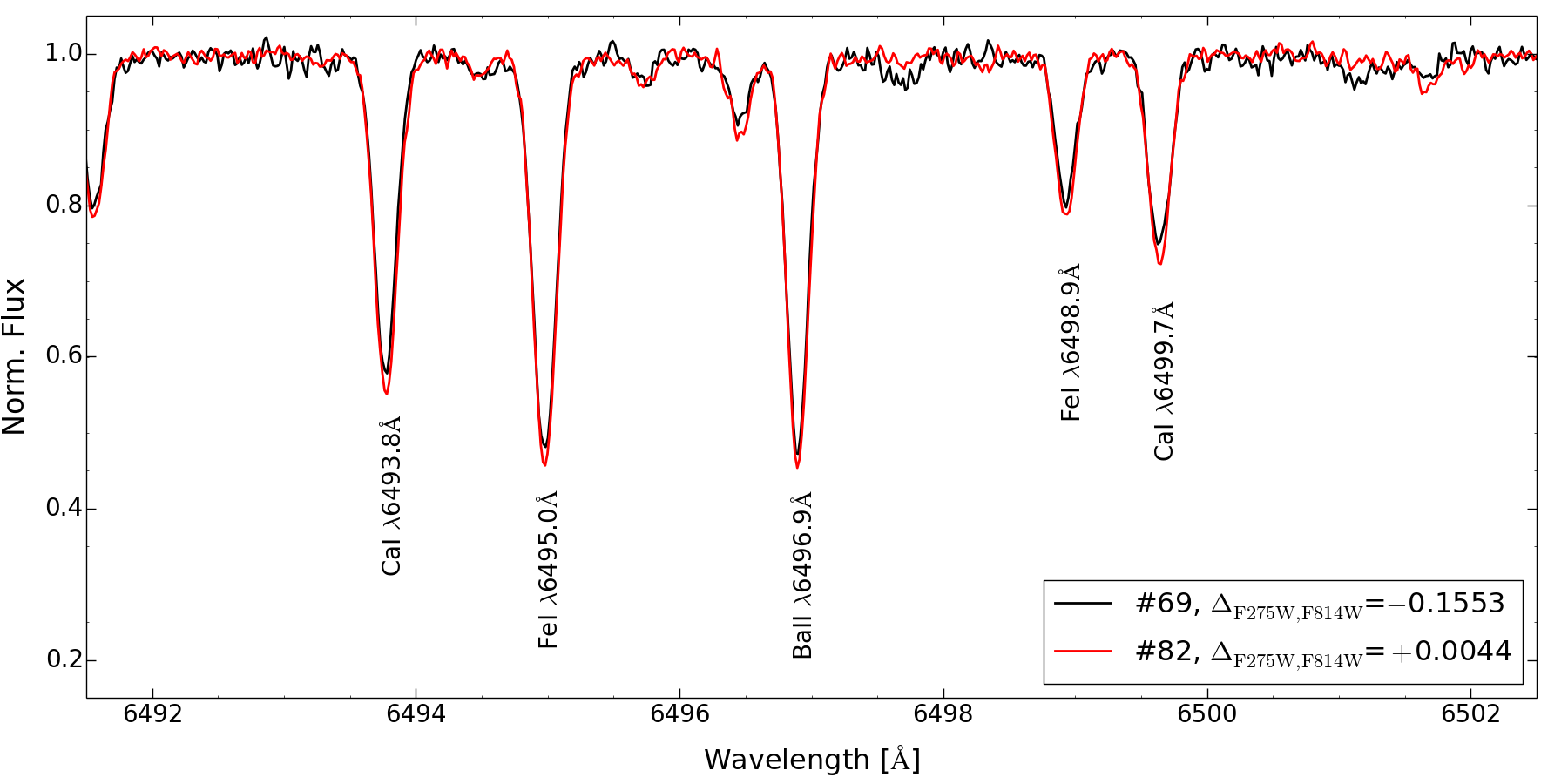}
      \caption{Examples of two portions of spectra including some Fe
        and Ca analysed spectral features, plus the Ba line
        $\lambda$6496.9\AA. The two represented spectra are for two
        stars with similar atmospheric parameters, namely \#69
        ((\teff,\logg,[A/H],\vmicro)=(4850~K/2.00/$-$1.55/1.48~\kmsec)),
      and \#82
      ((\teff,\logg,[A/H],\vmicro)=(4860~K/2.05/$-$1.44/1.49~\kmsec)),
    but different \x\ values. Overall, the spectral features of \#82,
    with higher \x, are deeper.}  
        \label{fig:spectra}
   \end{figure*}

A quick look at the spectra further supports this finding. As an
example, in Figure~\ref{fig:spectra} we show the portion of two spectra,
including some Fe and Ca analysed spectral features, for two stars
with different \x\ and similar atmospheric parameters. We note that
the two chosen stars also have very similar atmospheric parameters
from photometry. 
Overall, the spectral features of the star at higher \x\ look
consistent with higher metals.

As first suggested by \cite{Mil15, Mil17} a variation in the He
content intrinsic to the 1G stars is qualitatively able to account for
the \x\ spread. However, such a variation in He should be high, at a
level of $\Delta$Y$\sim$0.08-0.10, which is difficult to achieve
without any corresponding enhancement in other chemical species, such
as N and Na, and depletion in O \citep{Mar19}. As discussed in
Section~\ref{sec:results} our analysed stars in NGC\,3201 have
similar [O/Fe] and [Na/Fe] abundances. 

Assuming the working hypothesis that a He variation exists among our
stars, we would expect a difference in the structure of a model
atmosphere between a He-normal and a He-rich star
\citep{Stromgren82}. Such a difference would translate in a different
surface gravity, which is expected to be small ($<$0.10~dex). Looking
at the impact of a variation in \logg\ by 0.20~dex on the chemical
abundances, it is unlikely that such variation in \logg\ introduced by
a possible He variations are responsible for the abundance variations
we find. 
 
As widely discussed in \citet{Yong13}, a second effect of possible He
variations is that for a fixed mass fraction of metals (Z), a change
in the helium mass fraction (Y) will directly affect the hydrogen mass
fraction (X) such that the metal-to-hydrogen ratio, Z/X will change
with helium mass fraction since X + Y + Z = 1. Hence, if stars in a
globular cluster have a constant Z, an He-rich
star will appear to be slightly more metal-rich than an He-normal
star. 
In this context, by using spectra of excellent quality \citet{Yong13} found variations
at a level of a few hundredth of dex in the absolute chemical
abundances of giants in NGC\,6752, positively correlated with Na (a
tracer of He)\footnote{For NGC\,6752, high-precision photometry
infers a $\Delta$Y of the order of a few hundredth \citep{Mil18}, not enough to
entirely account for the observed abundance correlations derived in
\citet{Yong13}. Yong and co-workers suggest that a combination of He variations and
inhomogeneous chemical evolution in the protocluster environment could
account for the chemical abundance variations.}. 

For a fixed Z, in the case of a variation in Y of 0.10 (the one
predicted to account for the \x\ extention in NGC\,3201), [Z/H] would
change by $\sim$0.06~dex, which is slightly lower than the range we
find in the cluster.
However, as a {\it pure} He enhancement shifts stars towards lower \x\ values
\citep{Mil15, Mar19}, this second explanation can be ruled out for our
1G stars in NGC\,3201, as the stars that should be enhanced in He (at
lower \x), are metal-poorer. Clearly, they are located on the blue side
of the RGB, and of the ChM, reinforcing the idea that, at least the
bluest side of 1G in the ChM, is populated by binaries and/or evolved
blue stragglers. 

On the other hand, if we assume that all the metal-poorer and bluest stars are
indeed binaries, we expect that the sum of the spectra of the two
binary components is reflected in some variation in the line-flux/continuum
ratio. Depending on the brightness of the two stars, these effects
might not be negligible and can affect the derived atmospheric parameters. 
To qualitatively investigate this issue, we have simulated some
spectra with the SYNTHE routine in the ATLAS code
\citep{Kur09}. For this purpose, we considered a bright giant star with
(\teff/\logg/[A/H]/\vmicro)=(4650~K/1.60/$-$1.50/2.00~\kmsec), and two
less luminous stars, namely a fainter giant
(5000~K/2.53/$-$1.50/2.00~\kmsec), and a sub-giant
(6000~K/4.00/$-$1.50/2.00~\kmsec). By summing the flux of the bright
giant and the sub-giant, the combined spectrum is almost
identical to the giant spectrum. However, the sum of the two giants
results in an emerging spectrum which is consistent with either lower
overall metallicity (by 0.06~dex) or higher \teff\ (by 50~K). Binaries
where the two components are both giants, but with different
luminosities can potentially explain the lower metals in the bluest 1G
stars, however as far as we know it is unlikely to have as many giant
pairs as the bluer 1G stars \citep{APM12bin}.

\section{Discussion and conclusions}\label{sec:conclusions}

We have presented a high-resolution chemical abundance analysis of
eighteen stars of the GC NGC\,3201. The eighteen stars have been
selected all to belong to the 1G stellar population, as defined on the
ChM diagnostic tool in \citet{Mil17}. Although located on the 1G ChM
sequence, our targets span a large range in the \x\ axis.

Overall, we have found that the abundances relative to Fe of all the inferred
species are consistent with a uniform chemical
composition relative to Fe. Specifically, the stars have similar O and
Na, consistent 
with the 1G chemical abundances. This finding confirms previous results showing
that the elongation in \x\ of the 1G in GCs is not related to the
common light-elements (anti)-correlations \citep{Mar19, CZ19}.
The constancy in light elements makes it hard to believe that the stars
with lower \x\ have higher He abundances, as tentatively suggested by
\citet{Mil15}. 

The three stars with the lowest \x\ values in the sample are binary
candidates (see Figure~\ref{fig:LaVsRV}). Two of these stars have
higher RV r.m.s., over an
observation time of a few months, than the
bulk of our stars. One star (\#93) also shows extremely high $s$-process
element abundances. A high $s$-element abundance is associated to 
long-period, single-lined spectroscopic binary systems
\citep{McClure89}. The unseen companion was a low-mass AGB star that
transferred processed material to the surviving visible star. 
A third star shows higher $s$-element abundances, although not as
extreme as \#93, but its r.m.s. in RVs is similar to the bulk of
stars. This latter star has however a relatively short
observational time coverage, three days. 

These results strongly suggest that the stars with the lower \x\
values on the ChM, specifically those with \x$\lesssim -$0.18 in
NGC\,3201, are binaries. 
It is interesting to note that barium and CH-type stars form generally
via wind mass transfer in fairly wide binaries \citep{Jor98}. 
Such wide binaries are not likely to survive in a GC
because of dynamical interactions, which suggests that the two
$s$-rich stars either may not be in binaries systems today, or that 
formed via RLOF which can lead to tighter, shorter period binaries
more likely to survive in a GC.  

The observed distribution in \x\ is consistent with simulations of
non-interacting binaries
(presented in Section~\ref{sec:simu}) formed by two
stars belonging to the 1G population, which predict these objects on
lower \x\ values. The fact that 2G stars in the ChM do not show a large
spread in \x, as the 1G does, might support the idea of a predominance of 1G-1G
binaries in GCs \citep{Lucatello15}. 
Furthermore, we note that the BS population
accounts for the bluest stars in the elongated 1G population observed
on the ChM.

Excluding the three binary candidates, we found a small variation in
the overall metallicity in the remaining fifteen giants. [Fe/H] has a
range of the order of $\sim$0.1~dex, which is correlated to the \x\
value. The absolute abundances of the other analysed elements follow the same
variation of Fe, keeping constant the abundance ratios relative to
Fe. 
We exclude the possibility that this small vatiation in metals is introduced by a
change in helium (Y) and a consequent change in Z/X, as in this case
we would expect the opposite trend with the \x\ values.

We can interpret the observed small change in the overall metallicity
either as a hint of internal variations among 1G stars, or as an 
artifact of binarity.
In the first case, we have to account for an inhomogeneity in
the primordial cloud from which the GC formed.  In this context we
note that, when data of excellent quality are analysed, very small
internal variations in metals are also found in open clusters \citep{Liu16}.
On the other hand, our simulated spectra of binaries suggest that a
giant-giant pair (not identical) spectrum is consistent with higher
temperature and/or higher metallicity, as the one we find. Previous
work on the binary fraction of NGC\,3201 however does not support such
a high number of binaries as the one required to account for the
observed elongated 1G \citep{APM12bin}.

We conclude that binarity surely contributes to 
the elongation of the 1G in NGC\,3201.
On the other hand, only binaries with $q\gtrsim$0.8 can produce
a sizeble shift towards low \x\ values.
As an additional mechanism, a small inhomogeneity in metals can
account for some of the spread among 1G stars in the map. 
Helium variations, while able to theoretically produce 
the observed elongation along the 1G on the ChM, seem unlikely.

\acknowledgments
The authors warmly thank the anonymous referee for very insightful
discussion. 
This work has received funding from the European Research Council
(ERC) under the European Union's Horizon 2020 research innovation
programme (Grant Agreement ERC-StG 2016, No 716082 'GALFOR', PI:
Milone), and the European Union's Horizon 2020 research and innovation
programme under the Marie Sk{\l}odowska-Curie (Grant Agreement No
797100). 
APM and MT acknowledge support from MIUR through the FARE project R164RM93XW ``SEMPLICE''.
HJ acknowledges support by the Australian Research Council through the
Discovery Project DP150100862. 

\software{ATLAS code \citep{Kur09}, MOOG code \citep{moog}, EsoReflex
interface \citep{Ballester00}, ESO MOLECFIT tool \citep{smette14, kausch14}.}

\begin{deluxetable}{c cc c cc ccc c}
\tablewidth{10pt}
\tabletypesize{\scriptsize}
\tablecaption{Coordinates, Photometric Information (including the location
  on the ChM), Radial Velocities, with associated rms from the \#
  (number) exposures and maximum time between observations (Time interval).}\label{tab:targets}
\tablehead{
\colhead{ID} & RA  & DEC          & \colhead{$V$} & \colhead{\x} & \colhead{\y}           & \colhead{RV} & \colhead{r.m.s.$_{\mathrm {RV}}$} & \colhead{\#}  & \colhead{Time interval}   \\
             & J2000 & J2000      & \colhead{mag} & \colhead{mag}&\colhead{mag}          & \colhead{[\kmsec]} & \colhead{[\kmsec]} &   & }
\startdata
150 & 10:17:31.515 & $-$46:24:34.66      & 14.007 & $-$0.1118           & \phantom{$-$}0.0240   & 490.8 & 0.22 & 5  & 2 months\\
160 & 10:17:28.083 & $-$46:24:33.71      & 14.323 & $-$0.1432           & \phantom{$-$}0.0328   & 493.0 & 0.23 & 5  & 2 months\\
 32 & 10:17:33.746 & $-$46:25:46.50      & 14.577 & $-$0.1030           & \phantom{$-$}0.0005   & 498.6 & 0.14 & 5  & 2 months\\
 69 & 10:17:38.395 & $-$46:25:29.28      & 14.542 & $-$0.1553           & \phantom{$-$}0.0117   & 493.1 & 0.17 & 5  & 2 months\\
 58 & 10:17:40.771 & $-$46:24:44.87      & 13.649 & \phantom{$-$}0.0363 & $-$0.0014             & 498.1 & 0.19 & 5  & 2 months\\
 91 & 10:17:32.111 & $-$46:25:09.95      & 14.165 & $-$0.1662           & \phantom{$-$}0.0367   & 493.5 & 0.18 & 5  & 2 months\\
 40 & 10:17:29.962 & $-$46:25:30.57      & 13.821 & $-$0.0808           & \phantom{$-$}0.0038   & 496.8 & 0.21 & 5  & 2 months\\
147 & 10:17:33.391 & $-$46:24:25.34      & 14.794 & \phantom{$-$}0.0783 & \phantom{$-$}0.0030   & 493.6 & 0.20 & 12 & 8 months  \\
149 & 10:17:32.115 & $-$46:24:24.49      & 14.633 & $-$0.2749           & \phantom{$-$}0.0938   & 495.2 & 0.11 & 11 & 3 days    \\
14  & 10:17:41.299 & $-$46:25:43.50      & 14.674 & $-$0.0827           & \phantom{$-$}0.0077   & 496.0 & 0.20 & 12 & 7 months  \\
158 & 10:17:28.360 & $-$46:24:03.89      & 14.936 & $-$0.1326           & \phantom{$-$}0.0343   & 494.0 & 0.08 & 11 & 3 days    \\
15  & 10:17:41.161 & $-$46:25:33.81      & 14.684 & $-$0.0326           & \phantom{$-$}0.0042   & 497.7 & 0.09 & 11 & 3 days    \\
27  & 10:17:36.395 & $-$46:25:49.35      & 14.732 & \phantom{$-$}0.0026 & $-$0.0143             & 492.5 & 0.08 & 11 & 3 days    \\
33  & 10:17:33.501 & $-$46:25:30.71      & 15.397 & $-$0.1231           & \phantom{$-$}0.0600   & 492.0 & 0.24 & 23 & 8 months  \\
67  & 10:17:39.149 & $-$46:25:19.59      & 15.419 & $-$0.1994           & \phantom{$-$}0.0804   & 496.4 & 0.76 & 23 & 8 months  \\
82  & 10:17:35.886 & $-$46:25:21.01      & 14.685 & \phantom{$-$}0.0044 & $-$0.0197             & 492.6 & 0.19 & 12 & 7 months  \\
93  & 10:17:31.507 & $-$46:24:58.17      & 14.574 & $-$0.2174           & \phantom{$-$}0.0858   & 506.8 & 0.55 & 12 & 7 months  \\
98  & 10:17:30.035 & $-$46:25:00.41      & 15.350 & $-$0.1796           & \phantom{$-$}0.0809   & 493.9 & 0.26 & 23 & 8 months  \\
\hline
\enddata
\end{deluxetable}

\begin{deluxetable}{r cccc ccc ccc cc}
\tablewidth{10pt}
\tablecaption{Adopted Atmospheric Parameters, derived from
  spectroscopy, and corresponding Fe~\,{\sc i} and Fe~\,{\sc ii}
  abundances (with the associated $\sigma$ and number of spectral 
  lines \#). The last two columns list the \teff\ and \logg\
  values obtained from photometry.}\label{tab:atm}
\tablehead{
\colhead{ID} & \teff & \logg & [A/H] & \vmicro & log$\epsilon$(Fe{\sc i}) & $\sigma_{\mathrm {Fe{\sc I}}}$ & \#$_{\mathrm {Fe{\sc I}}}$ & log$\epsilon$(Fe{\sc ii}) & $\sigma_{\mathrm {Fe{\sc II}}}$ & \# & \teff$_{\mathrm {photometry}}$  & \logg$_{\mathrm {photometry}}$\\
             & (K) & (cgs)  & (dex) & (\kmsec) & dex &  &  & dex & & & (K) & (cgs)  }
\startdata
150   & 4680 & 1.60 & $-$1.55 & 1.53      & 5.95 & 0.09 & 128         & 6.01 & 0.075 & 12  &   4641 & 1.76 \\
160   & 4810 & 2.00 & $-$1.48 & 1.48      & 6.02 & 0.09 & 125         & 6.08 & 0.065 & 13  &   4729 & 1.94 \\
32    & 4830 & 2.00 & $-$1.46 & 1.44      & 6.04 & 0.11 & 131         & 6.10 & 0.075 & 12  &   4767 & 2.06 \\
69    & 4850 & 2.00 & $-$1.55 & 1.48      & 5.95 & 0.09 & 119         & 6.00 & 0.072 & 12  &   4781 & 2.05 \\
58    & 4650 & 1.60 & $-$1.44 & 1.64      & 6.06 & 0.09 & 133         & 6.14 & 0.051 & 13  &   4507 & 1.54 \\
91    & 4780 & 1.90 & $-$1.53 & 1.54      & 5.97 & 0.09 & 126         & 6.03 & 0.067 & 13  &   4726 & 1.87 \\
40    & 4700 & 1.70 & $-$1.48 & 1.60      & 6.01 & 0.10 & 132         & 6.09 & 0.059 & 13  &   4600 & 1.66 \\
147   & 4860 & 2.07 & $-$1.49 & 1.40      & 6.01 & 0.09 & 125         & 6.08 & 0.084 & 13  &   4648 & 2.08 \\
149   & 4920 & 2.15 & $-$1.47 & 1.46      & 6.03 & 0.10 & 122         & 6.11 & 0.066 & 13  &   4838 & 2.12 \\
14    & 4880 & 2.15 & $-$1.48 & 1.54      & 6.02 & 0.09 & 129         & 6.09 & 0.060 & 13  &   4760 & 2.09 \\
158   & 4910 & 2.20 & $-$1.56 & 1.42      & 5.94 & 0.11 & 117         & 6.00 & 0.060 & 12  &   4816 & 2.23 \\
15    & 4880 & 2.20 & $-$1.42 & 1.48      & 6.08 & 0.09 & 125         & 6.16 & 0.036 & 13  &   4762 & 2.10 \\
27    & 4880 & 2.10 & $-$1.39 & 1.48      & 6.11 & 0.09 & 122         & 6.18 & 0.070 & 13  &   4762 & 2.12 \\
33    & 4950 & 2.35 & $-$1.58 & 1.37      & 5.92 & 0.10 & 110         & 5.97 & 0.071 & 13  &   4901 & 2.45 \\
67    & 4990 & 2.36 & $-$1.68 & 1.41      & 5.82 & 0.10 & 111         & 5.88 & 0.090 & 13  &   4940 & 2.48 \\
82    & 4860 & 2.05 & $-$1.44 & 1.49      & 6.06 & 0.08 & 128         & 6.14 & 0.087 & 13  &   4780 & 2.11 \\
93    & 4950 & 2.33 & $-$1.46 & 1.55      & 6.03 & 0.09 & 124         & 6.11 & 0.055 & 13  &   4765 & 2.06 \\
98    & 5000 & 2.53 & $-$1.55 & 1.36      & 5.95 & 0.09 & 112         & 6.01 & 0.083 & 13  &   4922 & 2.44 \\\hline
\enddata
\end{deluxetable}
\newpage
\movetabledown=5.8cm
\begin{rotatetable}
\floattable
\begin{deluxetable}{ccccccccccccccccccccccccc}
\tablewidth{6pt}
\tabletypesize{\scriptsize}
\tablecaption{Analyzed Chemical Abundances from Li to Sc.\label{tab:LiToSc}}
\tablehead{
STAR & A(Li) &A(Li) &[O/Fe] & $\sigma$ & \# & [Na/Fe] & $\sigma$ & [Na/Fe] & \# & [Mg/Fe] & $\sigma$ & \# & [Al/Fe] & $\sigma$ & \# & [Si/Fe] & $\sigma$ & \# & [Ca/Fe] & $\sigma$ & \# & [Sc/Fe] & $\sigma$ & \# \\
     & LTE   &non-LTE&      &          &    &   LTE   &          & non-LTE &    &         &          &    &         &          &    &         &          &    &         &          &    &         &          &          
}
\startdata
150  & $<$0.37\tablenotemark{\tiny a}           & -- & 0.52 & 0.06 & 2 & $-$0.16         & 0.02 &$-$0.23 &4 & 0.34 & 0.01 & 2 & \phantom{$<$}$-$0.12 & 0.25 & 2 & 0.32 & 0.12 & 7 & 0.28 & 0.09 & 20 &$-$0.05           & 0.14 & 6 \\
160  & $<$0.46\tablenotemark{\tiny a}           & -- & 0.63 &   -- & 1 & $-$0.13         & 0.08 &$-$0.20 &4 & 0.23 & 0.01 & 2 & \phantom{$<$}$-$0.09 &   -- & 1 & 0.28 & 0.09 & 6 & 0.24 & 0.08 & 20 &\phantom{$-$}0.01 & 0.14 & 6 \\
32   & \phantom{$<$}0.94\tablenotemark{\tiny a} &1.05& 0.58 & 0.20 & 2 & $-$0.11         & 0.06 &$-$0.19 &3 & 0.26 & 0.04 & 2 & \phantom{$<$}$-$0.10 &   -- & 1 & 0.31 & 0.15 & 6 & 0.26 & 0.09 & 20 &\phantom{$-$}0.03 & 0.13 & 5 \\
69   & \phantom{$<$}0.74\tablenotemark{\tiny a} &0.84& 0.47 &   -- & 1 & $-$0.20         & 0.02 &$-$0.27 &3 & 0.27 & 0.01 & 2 & \phantom{$<$}$<$0.00 &   -- & - & 0.33 & 0.09 & 6 & 0.26 & 0.09 & 20 &\phantom{$-$}0.02 & 0.12 & 6 \\
58   & $<$0.31\tablenotemark{\tiny a}           & -- & 0.57 & 0.07 & 2 & $-$0.13         & 0.02 &$-$0.20 &4 & 0.28 & 0.04 & 2 & \phantom{$<$}$-$0.13 & 0.26 & 2 & 0.29 & 0.10 & 7 & 0.28 & 0.08 & 20 &$-$0.01           & 0.13 & 7 \\
91   & $<$0.37\tablenotemark{\tiny a}           & -- & 0.50 &   -- & 1 & $-$0.13         & 0.15 &$-$0.20 &4 & 0.28 & 0.02 & 2 & \phantom{$<$}$-$0.10 & 0.20 & 2 & 0.32 & 0.11 & 7 & 0.27 & 0.08 & 20 &\phantom{$-$}0.00 & 0.10 & 6 \\
40   & $<$0.27\tablenotemark{\tiny a}           & -- & 0.57 & 0.02 & 2 & $-$0.20         & 0.08 &$-$0.26 &4 & 0.28 & 0.02 & 2 & \phantom{$<$}$-$0.23 &   -- & 1 & 0.33 & 0.11 & 7 & 0.26 & 0.09 & 20 &\phantom{$-$}0.01 & 0.10 & 6 \\
147  & \phantom{$<$}0.91                        &1.02& 0.55 &   -- & 1 & $-$0.15         & 0.09 &$-$0.22 &4 & 0.29 & 0.02 & 2 & $<-$0.10             &   -- & - & 0.26 & 0.09 & 7 & 0.27 & 0.10 & 20 &$-$0.07           & 0.13 & 6 \\
149  & $<$0.50\tablenotemark{\tiny a}           & -- & 0.50 &   -- & 1 & $-$0.06         & 0.13 &$-$0.14 &4 & 0.27 & 0.02 & 2 & \phantom{$<$}$-$0.18 &   -- & 1 & 0.32 & 0.11 & 7 & 0.26 & 0.08 & 20 &$-$0.02           & 0.14 & 6 \\
14   & \phantom{$<$}0.94                        &1.04& 0.54 &   -- & 1 & $-$0.18         & 0.04 &$-$0.26 &4 & 0.27 & 0.03 & 2 & \phantom{$<$}$-$0.13 & 0.20 & 2 & 0.30 & 0.10 & 5 & 0.26 & 0.09 & 20 &\phantom{$-$}0.00 & 0.14 & 6 \\
158  & \phantom{$<$}0.95                        &1.04& 0.60 &   -- & 1 & $-$0.15         & 0.16 &$-$0.22 &4 & 0.25 & 0.02 & 2 & \phantom{$<$}$-$0.13 &   -- & 1 & 0.33 & 0.12 & 5 & 0.26 & 0.09 & 20 &\phantom{$-$}0.01 & 0.12 & 5 \\
15   & \phantom{$<$}0.93                        &1.03& 0.59 & 0.09 & 2 & $-$0.16         & 0.06 &$-$0.24 &4 & 0.26 & 0.02 & 2 & \phantom{$<$}$-$0.05 & 0.30 & 2 & 0.32 & 0.09 & 5 & 0.28 & 0.08 & 20 &\phantom{$-$}0.04 & 0.12 & 5 \\
27   & \phantom{$<$}0.97                        &1.07& 0.48 &   -- & 1 & $-$0.28         & 0.07 &$-$0.35 &4 & 0.20 & 0.01 & 2 & \phantom{$<$}$-$0.26 & 0.19 & 2 & 0.28 & 0.09 & 6 & 0.24 & 0.09 & 19 &$-$0.03           & 0.09 & 6 \\
33   & \phantom{$<$}0.80                        &0.88& 0.63 &   -- & 1 & $-$0.18         & 0.04 &$-$0.26 &2 & 0.24 & 0.01 & 2 & \phantom{$<$}$<$0.10 &   -- & - & 0.30 & 0.10 & 6 & 0.28 & 0.07 & 18 &$-$0.03           & 0.14 & 6 \\
67   & \phantom{$<$}0.87                        &0.94& 0.59 &   -- & 1 & $-$0.19         & 0.21 &$-$0.25 &4 & 0.27 & 0.01 & 2 & $<-$0.10             &   -- & - & 0.29 & 0.10 & 6 & 0.24 & 0.08 & 19 &$-$0.02           & 0.20 & 6 \\
82   & \phantom{$<$}0.80                        &0.91& 0.58 & 0.02 & 2 & $-$0.17         & 0.10 &$-$0.25 &4 & 0.22 & 0.01 & 2 & \phantom{$<$}$-$0.13 & 0.08 & 2 & 0.28 & 0.09 & 7 & 0.25 & 0.09 & 20 &$-$0.03           & 0.14 & 6 \\
93   & $<$0.44\tablenotemark{\tiny a}           & -- & 0.71 & 0.02 & 2 & $-$0.16         & 0.12 &$-$0.24 &4 & 0.28 & 0.00 & 2 & \phantom{$<$}$-$0.33 &   -- & 1 & 0.35 & 0.10 & 5 & 0.23 & 0.09 & 20 &\phantom{$-$}0.05 & 0.10 & 5 \\
98   & \phantom{$<$}0.90                        &0.97& 0.68 &   -- & 1 & $-$0.22         & 0.22 &$-$0.29 &4 & 0.22 & 0.00 & 2 & \phantom{$<$}$-$0.10 & 0.38 & 2 & 0.28 & 0.08 & 7 & 0.25 & 0.08 & 19 &$-$0.04           & 0.09 & 6 \\\hline
avg. &    0.89                                  &0.98& 0.57 &      &   & $-$0.16         &      &$-$0.23 &  & 0.26 &      &   & $-$0.15              &      &   & 0.31 &      &   & 0.26 &      &    &$-$0.01           &      & \\
$\pm$&    0.02                                  &0.02& 0.02 &      &   &\phantom{$-$}0.01&      &\phantom{$-$}0.01 && 0.01 && & \phantom{$-$}0.02    &      &   & 0.01 &      &   & 0.00 &      &    &\phantom{$-$}0.01 &      & \\
$\sigma$& 0.07                                  &0.08& 0.07 &      &   &\phantom{$-$}0.05&      &\phantom{$-$}0.05 && 0.03 && & \phantom{$-$}0.08    &      &   & 0.02 &      &   & 0.02 &      &    &\phantom{$-$}0.03 &      & \\
\enddata
\tablenotetext{\tiny a}{Star brighter than the RGB bump.}
\end{deluxetable}
\end{rotatetable}
\newpage
\movetabledown=5.cm
\begin{rotatetable}
\floattable
\begin{deluxetable}{cccccccccccccccccccccccccc}
\tablewidth{6pt}
\tabletypesize{\scriptsize}
\tablecaption{Analyzed Chemical Abundances from Ti to Ni.\label{tab:TiToNi}}
\tablehead{
STAR & [Ti/Fe]{\sc i} & $\sigma$ & \# & [Ti/Fe]{\sc ii} & $\sigma$ & \# & [V/Fe] & $\sigma$ & \# & [Cr/Fe]{\sc i} & $\sigma$ & \# & [Cr/Fe]{\sc ii} & $\sigma$ & \# & [Mn/Fe] & $\sigma$ & \# & [Co/Fe] & $\sigma$ & \# & [Ni/Fe] & $\sigma$ & \# }
\startdata
150 & 0.19 & 0.09 & 20 & 0.24 & 0.05 & 5 & $-$0.10 & 0.09 & 14 & $-$0.08 & 0.09 & 4 &\phantom{$-$}0.10 & 0.02 & 2 & $-$0.50 & 0.10 & 5 & $-$0.12          & 0.07 & 2 & $-$0.06          & 0.11 & 23 \\
160 & 0.25 & 0.12 & 20 & 0.21 & 0.09 & 5 & $-$0.06 & 0.11 & 13 & $-$0.08 & 0.11 & 4 &\phantom{$-$}0.10 & 0.03 & 2 & $-$0.46 & 0.07 & 5 & $-$0.23          & 0.00 & 2 & $-$0.03          & 0.12 & 23 \\
32  & 0.19 & 0.11 & 20 & 0.25 & 0.03 & 5 & $-$0.09 & 0.09 & 14 & $-$0.09 & 0.10 & 4 &\phantom{$-$}0.05 & 0.02 & 2 & $-$0.45 & 0.09 & 5 & $-$0.17          & 0.18 & 2 & $-$0.04          & 0.14 & 24 \\
69  & 0.25 & 0.12 & 18 & 0.29 & 0.05 & 5 & $-$0.02 & 0.11 &  9 & $-$0.15 & 0.07 & 4 &\phantom{$-$}0.16 & 0.02 & 2 & $-$0.49 & 0.08 & 5 & $-$0.09          & 0.01 & 2 & $-$0.04          & 0.13 & 20 \\
58  & 0.24 & 0.09 & 22 & 0.26 & 0.04 & 5 & $-$0.03 & 0.08 & 14 & $-$0.06 & 0.08 & 4 &\phantom{$-$}0.15 & 0.02 & 2 & $-$0.40 & 0.05 & 5 & $-$0.16          & 0.15 & 2 &\phantom{$-$}0.01 & 0.12 & 25 \\
91  & 0.19 & 0.12 & 20 & 0.24 & 0.06 & 5 & $-$0.11 & 0.12 & 13 & $-$0.15 & 0.12 & 4 &\phantom{$-$}0.07 & 0.02 & 2 & $-$0.50 & 0.12 & 5 & $-$0.10          & 0.15 & 2 & $-$0.04          & 0.11 & 22 \\
40  & 0.20 & 0.08 & 21 & 0.26 & 0.07 & 5 & $-$0.06 & 0.08 & 14 & $-$0.10 & 0.08 & 4 &\phantom{$-$}0.10 & 0.06 & 2 & $-$0.46 & 0.06 & 5 & $-$0.07          & 0.00 & 2 & $-$0.02          & 0.14 & 24 \\
147 & 0.24 & 0.09 & 19 & 0.27 & 0.05 & 5 & $-$0.10 & 0.07 & 13 & $-$0.08 & 0.10 & 4 &\phantom{$-$}0.10 & 0.12 & 2 & $-$0.48 & 0.12 & 5 & $-$0.23          & 0.09 & 2 & $-$0.05          & 0.12 & 21 \\
149 & 0.17 & 0.08 & 19 & 0.17 & 0.03 & 5 & $-$0.07 & 0.11 & 11 & $-$0.13 & 0.08 & 4 &\phantom{$-$}0.05 & 0.01 & 2 & $-$0.55 & 0.12 & 5 & $-$0.15          &   -- & 1 & $-$0.07          & 0.10 & 21 \\
14  & 0.23 & 0.07 & 22 & 0.27 & 0.08 & 5 & $-$0.02 & 0.06 &  9 & $-$0.05 & 0.11 & 4 &\phantom{$-$}0.18 & 0.02 & 2 & $-$0.44 & 0.06 & 5 & $-$0.20          &   -- & 1 & $-$0.01          & 0.10 & 24 \\
158 & 0.21 & 0.11 & 17 & 0.21 & 0.03 & 5 & $-$0.09 & 0.10 &  7 & $-$0.21 & 0.10 & 3 &\phantom{$-$}0.10 & 0.05 & 2 & $-$0.58 & 0.12 & 5 &\phantom{$-$}999  &   -- & - & $-$0.04          & 0.12 & 20 \\
15  & 0.20 & 0.06 & 21 & 0.25 & 0.05 & 5 & $-$0.02 & 0.09 & 11 & $-$0.11 & 0.06 & 4 &\phantom{$-$}0.11 & 0.04 & 2 & $-$0.50 & 0.10 & 5 & $-$0.10          & 0.05 & 2 & $-$0.02          & 0.12 & 24 \\
27  & 0.15 & 0.08 & 21 & 0.19 & 0.05 & 5 & $-$0.04 & 0.09 & 10 & $-$0.11 & 0.06 & 4 &          $-$0.01 & 0.14 & 2 & $-$0.50 & 0.08 & 5 & $-$0.14          & 0.03 & 2 & $-$0.04          & 0.11 & 24 \\
33  & 0.16 & 0.06 & 15 & 0.25 & 0.05 & 5 & $-$0.04 & 0.14 &  4 & $-$0.19 & 0.07 & 3 &\phantom{$-$}0.05 & 0.10 & 2 & $-$0.54 & 0.13 & 5 &\phantom{$-$}0.04 &   -- & 1 & $-$0.07          & 0.14 & 17 \\
67  & 0.21 & 0.07 & 14 & 0.18 & 0.11 & 5 & $-$0.07 & 0.08 &  6 & $-$0.25 & 0.12 & 2 &\phantom{$-$}0.17 & 0.08 & 2 & $-$0.50 & 0.14 & 5 &\phantom{$-$}999  &   -- & - & $-$0.08          & 0.12 & 15 \\
82  & 0.22 & 0.08 & 20 & 0.24 & 0.08 & 5 & $-$0.08 & 0.12 & 13 & $-$0.09 & 0.07 & 4 &\phantom{$-$}0.07 & 0.01 & 2 & $-$0.47 & 0.08 & 5 &$-$0.20           & 0.07 & 2 & $-$0.02          & 0.10 & 24 \\
93  & 0.24 & 0.09 & 19 & 0.32 & 0.07 & 5 & $-$0.05 & 0.07 &  9 & $-$0.05 & 0.15 & 4 &\phantom{$-$}0.17 & 0.02 & 2 & $-$0.47 & 0.10 & 5 &$-$0.19           & 0.19 & 2 & $-$0.03          & 0.10 & 22 \\
98  & 0.17 & 0.08 & 18 & 0.22 & 0.06 & 5 & $-$0.06 & 0.17 &  6 & $-$0.15 & 0.10 & 4 &\phantom{$-$}0.07 & 0.08 & 2 & $-$0.49 & 0.15 & 5 &\phantom{$-$}999  &   -- & - & $-$0.07          & 0.10 & 18 \\\hline
avg.& 0.21 &      &    & 0.24 &      &   & $-$0.06 &      &    & $-$0.12 &      &   & 0.10             &      &   & $-$0.49 &      &   & $-$0.14          &      &   & $-$0.04          &      &    \\
$\pm$&0.01 &      &    & 0.01 &      &   &\phantom{$-$}0.01&&  &\phantom{$-$}0.01&& & 0.01             &      &   &\phantom{$-$}0.01&& &\phantom{$-$}0.02 &      &   & \phantom{$-$}0.01&      &    \\
$\sigma$&0.03&    &    & 0.04 &      &   &\phantom{$-$}0.03&&  &\phantom{$-$}0.06&& & 0.05             &      &   &\phantom{$-$}0.04&& &\phantom{$-$}0.07 &      &   & \phantom{$-$}0.02&      &    \\
\enddata
\end{deluxetable}
\end{rotatetable}
\newpage
\movetabledown=3.cm
\begin{rotatetable}
\floattable
\begin{deluxetable}{ccccccccccccccccccccccccc}
\tablewidth{6pt}
\tabletypesize{\scriptsize}
\tablecaption{Analyzed Chemical Abundances from Cu to Eu.\label{tab:CuToEu}}
\tablehead{
STAR& [Cu/Fe] & [Zn/Fe]           &  [Y/Fe]            &$\sigma$&\#& [Zr/Fe]          &[Ba/Fe]&$\sigma$&\# &[La/Fe]& $\sigma$&\# &[Pr/Fe]&[Nd/Fe]&$\sigma$&\#&[Eu/Fe] }
\startdata                          
150 & $-$0.57 & \phantom{$-$}0.13 &  $-$0.24           & 0.07 & 3 &\phantom{$-$}0.04  & 0.20  & 0.04   & 3 & 0.17  & 0.09    & 7 & 0.23  & 0.10 & 0.03 & 2 & 0.45 \\
160 & $-$0.55 & \phantom{$-$}0.19 &  $-$0.18           & 0.08 & 3 &\phantom{$-$}0.11  & 0.33  & 0.05   & 3 & 0.27  & 0.11    & 5 & 0.33  & 0.25 & 0.02 & 2 & 0.60 \\
32  & $-$0.57 & \phantom{$-$}0.18 &  $-$0.21           & 0.07 & 2 &\phantom{$-$}0.03  & 0.32  & 0.03   & 3 & 0.25  & 0.07    & 6 & 0.32  & 0.28 & 0.06 & 2 & 0.56 \\
69  & $-$0.65 & \phantom{$-$}0.14 &  $-$0.20           & 0.12 & 2 &\phantom{$-$}0.04  & 0.21  & 0.04   & 3 & 0.25  & 0.05    & 6 & 0.33  & 0.20 & 0.05 & 2 & 0.47 \\
58  & $-$0.34 & \phantom{$-$}0.23 &  $-$0.24           & 0.07 & 3 &\phantom{$-$}0.16  & 0.25  & 0.02   & 3 & 0.24  & 0.08    & 7 & 0.30  & 0.27 & 0.08 & 2 & 0.56 \\
91  & $-$0.57 & \phantom{$-$}0.07 &  $-$0.17           & 0.15 & 3 &\phantom{$-$}0.02  & 0.21  & 0.01   & 3 & 0.14  & 0.05    & 6 & 0.32  & 0.13 & 0.04 & 2 & 0.28 \\
40  & $-$0.50 & \phantom{$-$}0.13 &  $-$0.27           & 0.04 & 3 &\phantom{$-$}0.13  & 0.19  & 0.06   & 3 & 0.18  & 0.06    & 6 & 0.18  & 0.13 & 0.07 & 2 & 0.43 \\
147 & $-$0.55 & \phantom{$-$}0.20 &  $-$0.27           & 0.03 & 3 &$-$0.01            & 0.27  & 0.05   & 3 & 0.27  & 0.06    & 6 & 0.40  & 0.17 & 0.09 & 2 & 0.54 \\
149 & $-$0.64 & \phantom{$-$}0.01 &  $-$0.19           & 0.04 & 3 &\phantom{$-$}0.15  & 0.68  & 0.04   & 3 & 0.47  & 0.07    & 6 & 0.45  & 0.42 & 0.01 & 2 & 0.45 \\
14  & $-$0.52 & \phantom{$-$}0.14 &  $-$0.15           & 0.17 & 3 &\phantom{$-$}0.02  & 0.30  & 0.07   & 3 & 0.29  & 0.06    & 6 & 0.20  & 0.19 & 0.05 & 2 & 0.55 \\
158 & $-$0.70 & $-$0.03           &  $-$0.26           & 0.01 & 2 &\phantom{$-$}0.09  & 0.22  & 0.08   & 3 & 0.20  & 0.13    & 2 &$<$0.55& 0.09 & 0.00 & 2 &$<$0.70 \\
15  & $-$0.61 & \phantom{$-$}0.11 &  $-$0.19           & 0.10 & 3 &\phantom{$-$}0.20  & 0.29  & 0.07   & 3 & 0.28  & 0.09    & 4 & 0.34  & 0.17 & 0.00 & 2 & 0.50 \\
27  & $-$0.62 & \phantom{$-$}0.04 &  $-$0.32           & 0.12 & 3 &\phantom{$-$}0.01  & 0.19  & 0.04   & 3 & 0.17  & 0.12    & 6 & 0.35  & 0.08 & 0.04 & 2 & 0.45 \\
33  & $-$0.67 & \phantom{$-$}0.21 &  $-$0.23           & 0.10 & 2 &\phantom{$-$}0.22  & 0.23  & 0.04   & 3 & 0.19  & 0.01    & 2 &$<$0.70& 0.21 & 0.00 & 2 & 0.41 \\
67  & $-$0.75 & $-$0.01           &  $-$0.33           & 0.06 & 2 &$-$0.05            & 0.15  & 0.10   & 3 & 0.19  & 0.03    & 2 &$<$0.80& 0.10 & 0.01 & 2 & 0.45 \\
82  & $-$0.52 & \phantom{$-$}0.12 &  $-$0.21           & 0.07 & 3 &\phantom{$-$}0.08  & 0.30  & 0.10   & 3 & 0.26  & 0.08    & 7 & 0.30  & 0.19 & 0.07 & 2 & 0.48 \\
93  & $-$0.59 & \phantom{$-$}0.12 &  \phantom{$-$}0.11 & 0.07 & 3 &\phantom{$-$}0.33  & 1.16  & 0.04   & 3 & 0.90  & 0.07    & 7 & 0.80  & 0.91 & 0.03 & 2 & 0.65 \\
98  & $-$0.60 & \phantom{$-$}0.03 &  $-$0.30           & 0.07 & 2 &\phantom{$-$}  --  & 0.23  & 0.08   & 3 & 0.26  & 0.14    & 2 &$<$0.50& 0.10 & 0.01 & 2 & 0.35\\\hline
avg.& $-$0.58 & \phantom{$-$}0.11 &  $-$0.21           &      &   &\phantom{$-$}0.09  & 0.32  &        &   & 0.28  &         &   &\phantom{$<$}0.35 &0.22&      &   & 0.48 \\
$\pm$&\phantom{$-$}0.02&\phantom{$-$}0.02& \phantom{$-$}0.02  &      &   &\phantom{$-$}0.02  & 0.06  &        &   & 0.04  &         &   &\phantom{$<$}0.04 &0.05&      &   & 0.02 \\
$\sigma$&\phantom{$-$}0.09&\phantom{$-$}0.08  & \phantom{$-$}0.10  &      &   &\phantom{$-$}0.10  & 0.24  &        &   & 0.17  &         &   &\phantom{$<$}0.15 &0.19&      &   & 0.09 \\
\enddata
\end{deluxetable}
\end{rotatetable}
\clearpage


\startlongtable
\begin{deluxetable}{lccccccc}
\tablewidth{10pt}
\tablecaption{Sensitivity of the derived abundances to the
  uncertainties in atmospheric parameters
  (\teff/\logg/\vmicro/[A/H]=$\pm$50~K/$\pm$0.20/$\pm$0.20~\kmsec/$\pm$0.10), 
  and uncertainties due to the errors in the EWs measurements or in
  the $\chi$-square fitting procedure. For reference, we also list the
  variations due to a change in \teff\ by $\pm$100~K. We reported the total internal
  uncertainty ($\sigma_{\rm total}$) obtained by the quadratic sum of
  all the contributers to the error.}\label{tab:err} 
\tablehead{
      &\colhead{$\Delta$\teff} & \colhead{$\Delta$\teff}&\colhead{$\Delta$\logg}&\colhead{$\Delta$\vmicro} &\colhead{$\Delta$[A/H]} &\colhead{$\sigma_{\rm EWs/fit}$}&\colhead{$\sigma_{\rm total}$}\\  
      &\colhead{$\pm$100~K}    & \colhead{$\pm$50~K}    & \colhead{$\pm$0.20}   &\colhead{$\pm$0.20~\kmsec}& \colhead{$\pm$0.10~dex}&                             &                   }
\startdata
$\rm {A(Li)}$            & $\pm$0.15   &$\pm$0.10    & $\pm$0.02  & $\pm$0.03  & $\pm$0.00 & $\pm$0.10       & $\pm$0.15  \\
$\rm {[O/Fe]}$           & $\pm$0.03   &$\pm$0.20    & $\pm$0.08  & $\pm$0.00  & $\pm$0.03 & $\pm$0.01       & $\pm$0.22  \\
$\rm {[Na/Fe]}$          & $\mp$0.06   &$\pm$0.05    & $\mp$0.02  & $\pm$0.04  & $\mp$0.01 & $\mp$0.02       & $\pm$0.07 \\
$\rm {[Mg/Fe]}$          & $\mp$0.06   &$\pm$0.02    & $\mp$0.04  & $\mp$0.03  & $\mp$0.02 & $\mp$0.02       & $\pm$0.08   \\
$\rm {[Al/Fe]}$          & $\pm$0.05   &$\pm$0.27    & $\mp$0.01  & $\pm$0.00  & $\mp$0.01 & $\pm$0.02       & $\pm$0.27 \\
$\rm {[Si/Fe]}$          & $\mp$0.08   &$\pm$0.02    & $\pm$0.03  & $\pm$0.05  & $\pm$0.03 & $\mp$0.05       & $\pm$0.08    \\
$\rm {[Ca/Fe]}$          & $\mp$0.03   &$\pm$0.01    & $\mp$0.01  & $\mp$0.01  & $\pm$0.00 & $\mp$0.01       & $\pm$0.02 \\
$\rm {[Sc/Fe]}$\,{\sc ii}& $\pm$0.09   &$\pm$0.02    & $\pm$0.04  & $\pm$0.07  & $\pm$0.05 & $\pm$0.03       & $\pm$0.10    \\
$\rm {[Ti/Fe]}$\,{\sc i} & $\pm$0.03   &$\pm$0.01    & $\mp$0.01  & $\pm$0.03  & $\pm$0.00 & $\pm$0.02       & $\pm$0.04    \\
$\rm {[Ti/Fe]}$\,{\sc ii}& $\pm$0.08   &$\pm$0.02    & $\pm$0.04  & $\pm$0.05  & $\pm$0.05 & $\pm$0.02       & $\pm$0.09    \\
$\rm {[V/Fe]}$           & $\pm$0.07   &$\pm$0.03    & $\pm$0.02  & $\pm$0.07  & $\pm$0.03 & $\pm$0.02       & $\pm$0.09    \\
$\rm {[Cr/Fe]}$\,{\sc i} & $\mp$0.03   &$\pm$0.03    & $\mp$0.06  & $\mp$0.04  & $\mp$0.05 & $\mp$0.01       & $\pm$0.092    \\
$\rm {[Cr/Fe]}$\,{\sc ii}& $\pm$0.02   &$\pm$0.05    & $\pm$0.01  & $\pm$0.04  & $\pm$0.01 & $\pm$0.00       & $\pm$0.07    \\
$\rm {[Mn/Fe]}$          & $\pm$0.03   &$\pm$0.07    & $\mp$0.02  & $\pm$0.04  & $\mp$0.01 & $\pm$0.01       & $\pm$0.08  \\
$\rm {[Fe/H]}$\,{\sc i}  & $\pm$0.13   &$\pm$0.00    & $\pm$0.01  & $\mp$0.04  & $\pm$0.01 & $\pm$0.06       & $\pm$0.07   \\
$\rm {[Fe/H]}$\,{\sc ii} & $\mp$0.07   &$\pm$0.01    & $\pm$0.08  & $\mp$0.07  & $\pm$0.01 & $\mp$0.02       & $\pm$0.11   \\
$\rm {[Co/Fe]}$          & $\pm$0.04   &$\pm$0.06    & $\pm$0.04  & $\pm$0.09  & $\pm$0.05 & $\pm$0.00       & $\pm$0.13    \\
$\rm {[Ni/Fe]}$          & $\mp$0.01   &$\pm$0.01    & $\pm$0.01  & $\pm$0.04  & $\pm$0.01 & $\mp$0.01       & $\pm$0.04    \\
$\rm {[Cu/Fe]}$          & $\pm$0.14   &$\pm$0.10    & $\mp$0.00  & $\mp$0.03  & $\mp$0.01 & $\pm$0.07       & $\pm$0.13    \\
$\rm {[Zn/Fe]}$          & $\mp$0.14   &$\pm$0.05    & $\pm$0.05  & $\mp$0.04  & $\pm$0.01 & $\mp$0.07       & $\pm$0.11    \\
$\rm {[Y/Fe]}$\,{\sc ii} & $\pm$0.07   &$\pm$0.05    & $\mp$0.01  & $\pm$0.01  & $\pm$0.01 & $\pm$0.03       & $\pm$0.06    \\
$\rm {[Zr/Fe]}$\,{\sc ii}& $\mp$0.04   &$\pm$0.15    & $\pm$0.10  & $\mp$0.00  & $\pm$0.05 & $\mp$0.02       & $\pm$0.19    \\
$\rm {[Ba/Fe]}$\,{\sc ii}& $\pm$0.03   &$\pm$0.10    & $\pm$0.07  & $\mp$0.15  & $\pm$0.03 & $\pm$0.01       & $\pm$0.20    \\
$\rm {[La/Fe]}$\,{\sc ii}& $\pm$0.03   &$\pm$0.07    & $\pm$0.07  & $\mp$0.01  & $\pm$0.03 & $\pm$0.01       & $\pm$0.10    \\
$\rm {[Pr/Fe]}$\,{\sc ii}& $\pm$0.04   &$\pm$0.15    & $\pm$0.07  & $\pm$0.00  & $\pm$0.05 & $\pm$0.01       & $\pm$0.17    \\
$\rm {[Nd/Fe]}$\,{\sc ii}& $\pm$0.14   &$\pm$0.06    & $\pm$0.06  & $\pm$0.09  & $\pm$0.07 & $\pm$0.05       & $\pm$0.15    \\
$\rm {[Eu/Fe]}$\,{\sc ii}& $\pm$0.00   &$\pm$0.15    & $\pm$0.08  & $\pm$0.00  & $\pm$0.04 & $\mp$0.00       & $\pm$0.17   \\
\hline
\enddata
\end{deluxetable}

\begin{deluxetable}{lcc}
\tablewidth{10pt}
\tablecaption{Spearman correlation coefficient ($r$) for each
log$\epsilon$(X) abundance as a function of \x, and probability that
the corresponding slope
is due to randomness ($f$).}\label{tab:spearman}
\tablehead{
\colhead{Element}         &\colhead{$r$} &\colhead{$f$ (\%)}   }
\startdata
Fe           &  0.68     &  2.2  \\
Li$_{\rm {non-LTE}}$& 0.35 &  27.6 \\
O            &  0.38     &  33.1 \\
Na$_{\rm {non-LTE}}$& 0.56 &  3.1  \\
Mg           &  0.70     &  3.7  \\
Al           &  0.37     &  37.6 \\
Si           &  0.51     &  9.9  \\
Ca           &  0.75     &  0.0  \\
Sc\,{\sc ii} &  0.47     &  8.2  \\
Ti\,{\sc i}  &  0.68     &  0.0  \\
Ti\,{\sc ii} &  0.75     &  2.2  \\
V            &  0.57     &  3.5  \\
Cr\,{\sc i}  &  0.68     &  0.3  \\
Cr\,{\sc ii} &  0.61     &  0.5  \\
Mn           &  0.63     &  0.2  \\
Co           &  0.09     & 40.0  \\
Ni           &  0.67     &  0.0  \\
Cu           &  0.66     &  1.0  \\
Zn           &  0.65     &  1.0  \\
Y\,{\sc ii}  &  0.37     &  4.8  \\
Zr\,{\sc ii} &  0.38     & 25.2  \\
Ba\,{\sc ii} &  0.55     & 18.7  \\
La\,{\sc ii} &  0.60     &  2.9  \\
Pr\,{\sc ii} &  0.57     & 15.5  \\
Nd\,{\sc ii} &  0.52     & 10.1  \\
Eu\,{\sc ii} &  0.63     &  4.1  \\
\enddata
\end{deluxetable}

\begin{table}
\caption{Atomic data and equivalent widths for program stars}\label{tab:linelist}
\begin{tabular}{c c c r c}
\hline
Wavelength [\AA] & Species & E.P. [eV] & log(gf) & EW [m\AA]\\
\hline
\multicolumn{5}{c}{150} \\
6806.860 & 26.0 & 2.730  & $-$3.140 &  21.9  \\
4885.430 & 26.0 & 3.880  & $-$1.150 &  46.9  \\
4917.240 & 26.0 & 4.190  & $-$1.270 &  32.5  \\
6581.220 & 26.0 & 1.490  & $-$4.680 &  14.7  \\
6592.910 & 26.0 & 2.730  & $-$1.490 &  99.2  \\
6593.880 & 26.0 & 2.430  & $-$2.420 &  76.6  \\
6608.040 & 26.0 & 2.280  & $-$3.960 &  10.8  \\
6609.120 & 26.0 & 2.560  & $-$2.690 &  52.9  \\
6627.560 & 26.0 & 4.550  & $-$1.500 &  6.6   \\
\hline
\end{tabular}
Notes:
Only a portion of this table is shown here to demonstrate its form and
content. A machine-readable version of the full table will be available.
\end{table}

\end{document}